\documentclass[aps,prl,preprint,groupedaddress]{revtex4-1}

\usepackage{graphicx}
\usepackage{colortbl}
\usepackage{xcolor}
\usepackage{multirow}
\usepackage{enumitem}
\usepackage{float}
\begin{document}

\title{Spin- and orbital-Hall effect in cyclic group symmetric metasurface}%

\author{Yeon Ui Lee$^{1}$, Igor Ozerov$^2$, Frederic Bedu$^2$, Ji Su Kim$^1$, Frederic Fages$^2$, and Jeong Weon Wu$^1$}%
\email{jwwu@ewha.ac.kr}
\affiliation{$^{1}$Department of Physics, Quantum Metamaterial Research Center, Ewha Womans University, Seoul 03760, Republic of Korea,
$^2$Aix Marseille Univ, CNRS, CINaM, Marseille, France}
\date{\today}%

\maketitle
\section*{abstract}
Light possesses both spin and orbital angular momentum (AM). While spin AM is determined by helicity of circular-polarization, orbital AM is characterized by topological charge of vortex beam. Interaction of AM with optical beam orbit leads to optical spin Hall or orbital Hall effect, exhibited as spin-dependent or topological charge-dependent transverse shift of optical beam. Conservation of AM enables spin-to-orbital AM conversion, where circular-polarized Gaussian beam is converted to opposite-helicity circular-polarized vortex beam with topological charge $\pm 2$, an example of controlling spatial beam profiling by spin flip. However, the resultant vortex beam has the beam center of gravity unchanged, the same as that of incident Gaussian beam, meaning a null transverse shift. Here we introduce a cyclic group symmetric metasurface to demonstrate generation of vortex beam exhibiting spin-dependent transverse shift, namely, spin- and orbital-Hall effect, attributed to an alteration of dynamical phase of scattered beam according to the order $n$ of cyclic group while keeping geometric phase constant. Capability of spin-controlled spatial beam profiling with a transverse shift via spin- and orbital-Hall effect has important implications for spatial demultiplexing in optical communication utilizing orbital AM mode division multiplexing as well as for optical vortex tweezer and signal processing involving vortex beams.

\section*{Introduction}
Circularly polarized (CP) light possesses spin angular momentum (SAM) with a helicity defined as its projection to linear momentum of propagating optical wave. In addition to SAM, optical wave can be rendered to possess orbital angular momentum (OAM) by constructing a vortex beam having helical wavefront.\cite{allen1992orbital} Both SAM and OAM are utilized to carry informations in optical communication via  polarization-division or OAM-division multiplexing. Differently from 2 modes of SAM with spins of $\pm 1$, the number of modes of OAM is not restricted since the topological charge of vortex beam is determined by twisting number of the helical wavefront.  Owing to the availability of a large number of modes, vortex beams have been widely employed in applications to increase the signal channels for optical communications.\cite{bozinovic2013terabit,ren2016chip,yang2016wavelength}
In those applications it is necessary to be able to manipulate and control the intensity and phase distributions of vortex beam.
Anisotropic planar structures such as metasurface are important optical elements for such applications when SAM is utilized for shaping vortex beam, since interaction of SAM and OAM with optical beam orbit is enhanced.\cite{bliokh2015spin}

There are several well-known means to create vortex beams, for example, a pair of cylindrical lenses, spiral phase plate, fork-shaped grating, and spatial light modulator as well as photonic integrated devices.\cite{allen1999iv,padgett2004light,cai2012integrated,li2015orbital}
Another important means to create vortex beams is to make use of geometric phase by constructing space-variant Pancharatnam-Berry phase optical elements including
subwavelength diffraction grating\cite{biener2002formation} and liquid crystal q-plates in spin-to-orbital AM conversion (SOC) process.\cite{marrucci2006optical}
Related to metasurface, a high-efficiency SOC has been demonstrated to generate vortex beams with high topological charges in the visible wavelength based on dielectric metasurfaces.\cite{devlin2016spin}
Generation of vortex beam by SOC is an example of spin-controlled spatial beam profiling since an incident Gaussian beam converts to a vortex beam with donut-shaped spatial beam profile by spin flip. In case of a rotational symmetric Pancharatnam-Berry phase optical element, SOC-generated vortex beams with opposite signs of topological charges are  degenerate in spatial beam profiles, which results from conservation of angular momentum consisting of spin AM and orbital AM. In other words, the beam centers of gravity of vortex beams are the same at that of the incident Gaussian beam.


Now we address the question how the degeneracy of spatial beam profiles of SOC-generated vortex beams can be lifted in SOC to provide spin-controlled spatial beam profiling such that the beam centers of gravity of vortex beams with opposite signs of topological charges are spatially separated by transverse-shifting in opposite directions.
Optical Hall effect is one means to achieve spatial separation by use of spin-dependent transverse shift originating from SAM- or OAM-orbit interaction, 
which is also technically important when spatially demultiplexing OAM-mode division multiplexed superposition states.

Here, we introduce a cyclic group symmetric metasurface (CGSM) to generate vortex beam exhibiting a spin- and orbital-AM dependent transverse shift by identifying SOC in terms of geometric phase acquired by an optical beam in cross-polarization scattering from ring structure of nano-antennas.
By designing CGSMs belonging to the cyclic group $C_n$, dynamical phase of cross-polarization scattered beam is altered according to the order $n$ of cyclic group while keeping geometric phase constant. When $n$-fold rotational symmetry of azimuthal dynamical phase gradient is an odd integer $n$, there results a spin- and orbital-Hall effect.\cite{bliokh2010angular}

\section{Geometric and dynamical phases in optical Hall effect}
\label{sec:geometirc}
Intrinsic optical spin or orbital Hall effect refers to spin-dependent or vortex-charge dependent transverse shift, a consequence of the coupling between SAM or OAM with optical beam orbit originating from transversality of electromagnetic wave.\cite{shitrit2013spin,cardano2015spin}. In a general framework of physical description, it is an effect of fast degree of freedom (SAM or OAM) on slow degree of freedom (optical beam orbit), described by momentum-space Lorentz equation of motion in the presence of a Berry curvature of topological monopole. \cite{Berry-BookChapter,onoda2004hall,bliokh2008geometrodynamics}
 %
%

\begin{figure*}[h]
\begin{center}
\includegraphics[width=12cm]{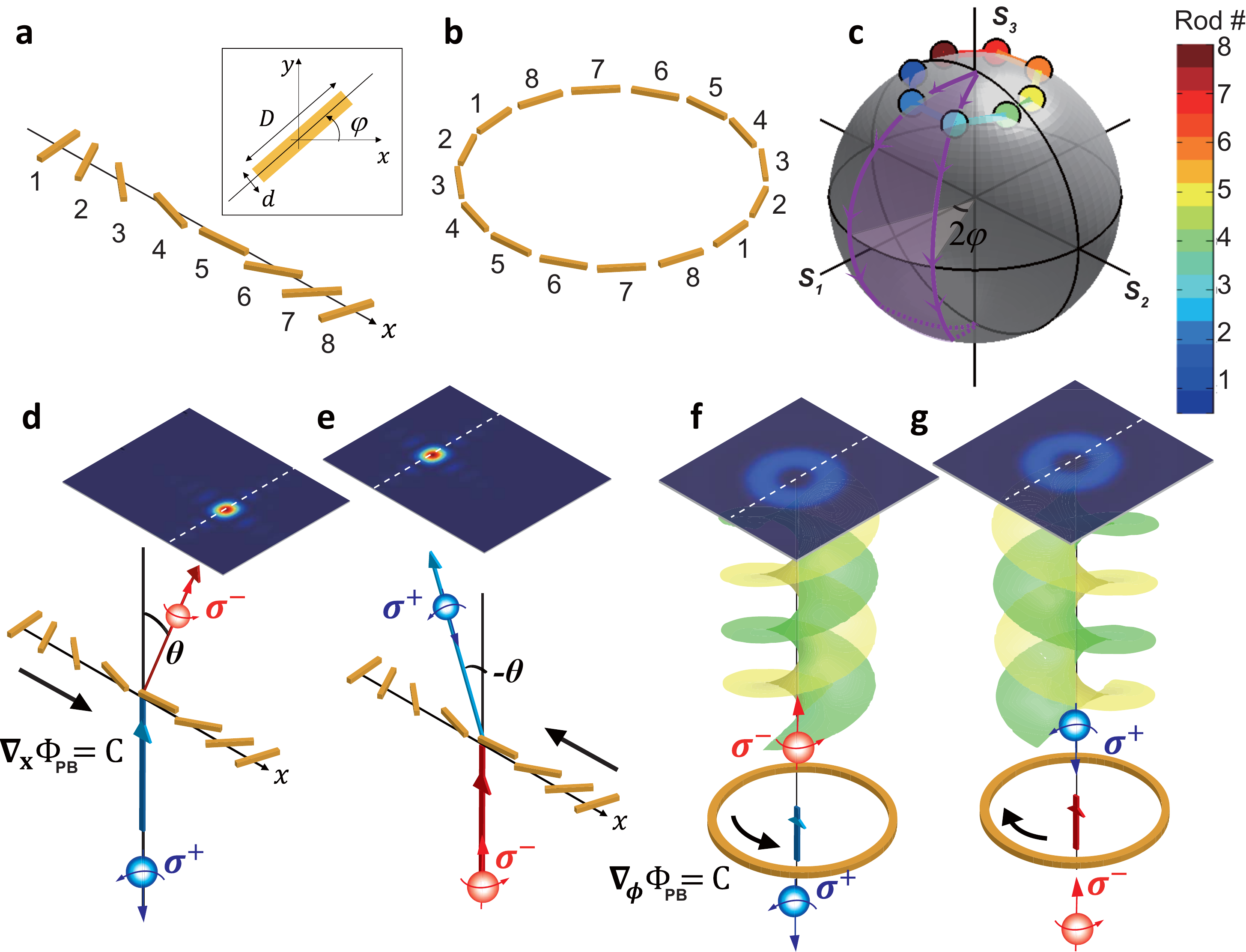}
\end{center}
\caption{\label{constantPB}\textbf{Array of nano-antennas with a constant PB phase-gradient and Poincar\'e sphere plot of Stokes parameters}. Inset figure shows a schematic of rotated nano-antenna with length $D$ and width $d$ making an angle $\varphi$ with $x$-axis. \textbf{a}, Linear array of rotating nano-antennas from $1$ to $8$ \textbf{b}, circular array of nano-antennas from $1$ to $8$ in the range of $0 \leq \phi  \leq \pi $ and from $1$ to $8$ in the range of $\pi \leq \phi  \leq 2\pi $ \textbf{c}, Poincar\'e sphere plot of Stokes parameters of optical beams scattered from nano-antennas of (a) and (b) for LCP incident beam. Color code corresponds to the nano-antenna number. \textbf{d,e}, spin-dependent deflection of cross-polarization scattered optical beam shown up as extrinsic spin Hall effect. $+x$-direction deflection of RCP ($\sigma_{-}$) beam for LCP ($\sigma_{+}$) incidence beam in (d) and the opposite deflection in (e).
\textbf{f,g}, generation of $l=+2$ vortex scattered beam of RCP ($\sigma_{-}$) for Gaussian incidence beam of LCP ($\sigma_{+}$) in (f) and the vortex with sign of topological charge reversed in (g).
}
\end{figure*}

%
Spin-dependent transverse shift of optical beam is also achieved by geometric Pancharatnam-Berry (PB) phase in a linear array of nano-antennas, called extrinsic optical spin Hall effect, to distinguish from the transverse shift originating from topological monopole.
The amount of shift is related to PB phase as shown in \textbf{Fig.~\ref{constantPB} a, d, e}.\cite{luo2015photonic}

In a circular array of nano-antennas with uniform width and length, even though not exhibiting spin-dependent transverse shift, spin-dependent PB phase is utilized to generate vortex beam by SOC in \textbf{Fig.~\ref{constantPB} b, f, g}.\cite{biener2002formation,marrucci2006optical, brasselet2013topological,karimi2014generating,osorio2015k}
Polarization states in Poincar\'e sphere go through a circular trajectory twice providing the solid angle $\pm8\pi$ corresponding to  geometric phase  $\pm4\pi$, generating vortex beam of $l=\pm 2$, as shown in \textbf{Fig.~\ref{constantPB} c}. 
Optical wave  scattered from the circular array acquires a phase factor, entirely geometric phase coming from a circular closed path $\textbf{C}$ in Stokes parameter space, leading to $\oint d\gamma(\textbf{C}) = \pm 4\pi$ with a constant azimuthal PB phase gradient $\nabla_\phi \Phi_{PB} = \pm 2$.
Therefore, the intensity profiles of $I_{-+}(r,\phi)=|{E}_{-+}|^2$ and $I_{+-}(r,\phi)=|{E}_{+-}|^2$ are degenerate even though helical phases are opposite for $l=\pm 2$, where $\{+-\}$ stands for left-circular polarization (LCP, $\sigma_{+}$) scattering / right-circular polarization (RCP, $\sigma_{-}$) incidence and $\{-+\}$ vice-versa.
In fact, ${E}_{-+}(r,\phi) = J_2(k_0r)e^{-{k_0}^2r^2}\exp(i2\phi)$ and ${E}_{+-}(r,\phi) = J_{-2}(k_0r)e^{-{k_0}^2r^2}\exp(-i2\phi)$, where a constant $\nabla_\phi \Phi_{PB}$ leads to $\phi$-independent amplitude $J_{\pm 2}(r)e^{-{k_0}^2r^2}$ from the fact that each two neighboring nano-antennas provide an equal amount of PB phase during one full cycle, owing to rotational invariance of circular array of nano-antennas.

It has been reported that a spin-dependent transverse shift of vortex beam in SOC is observed by lifting degeneracy in OAM though rotational symmetry breaking without
clarifying the underlying role of azimuthal phase gradient in giving rise to extrinsic spin- and orbital-Hall effect.\cite{shitrit2011optical,shitrit2013rashba,liu2015photonic,ling2015giant}
Now we examine how the azimuthal phase gradient is closely related to the existence of extrinsic spin- and orbital-Hall effect.

\begin{figure*}[h]
\begin{center}
\includegraphics[width=12cm]{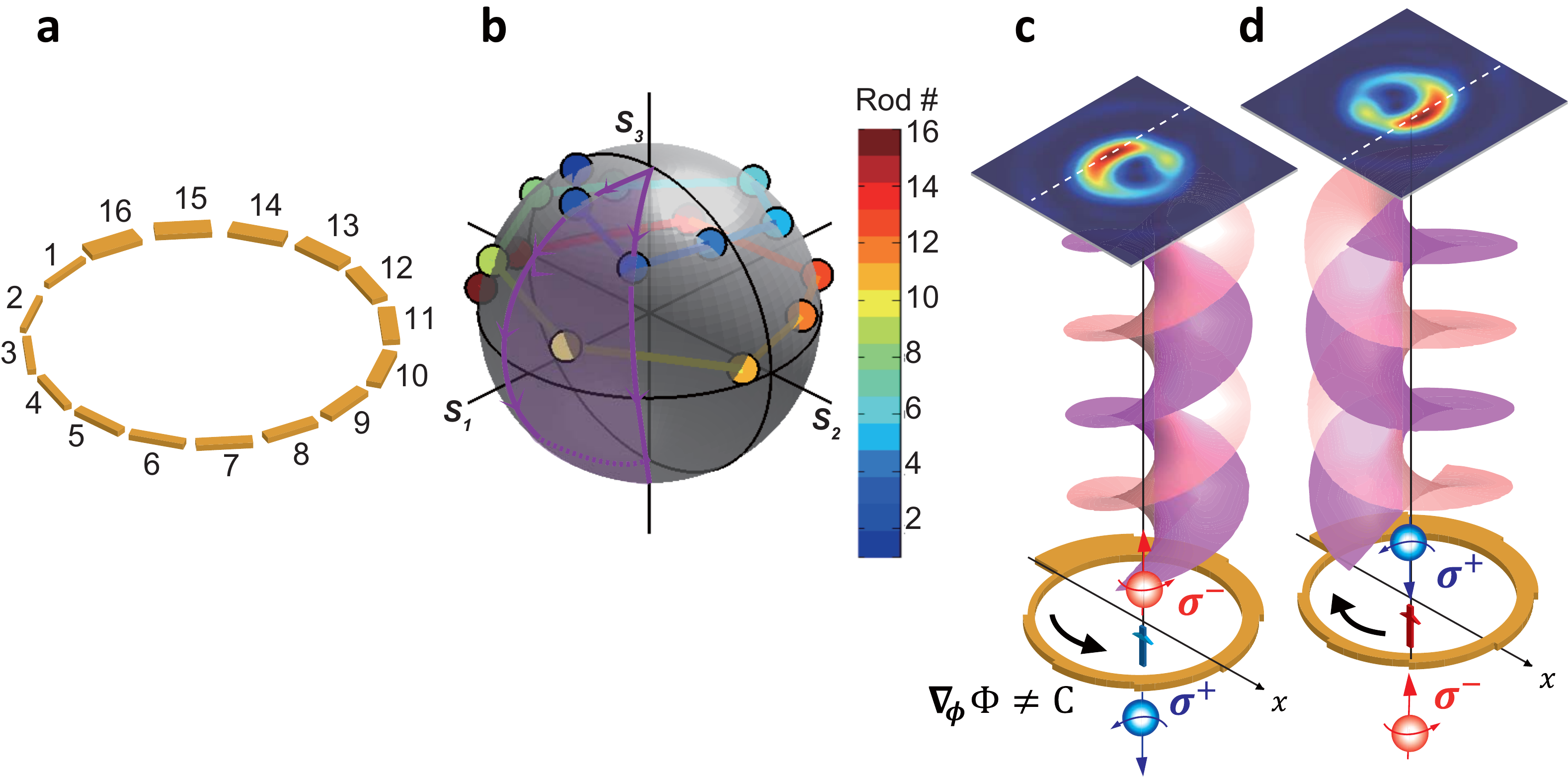}
\end{center}
\caption{\label{nonconstantPB}\textbf{Array of nano-antennas with a non-constant PB phase-gradient and Poincar\'e sphere plot of Stokes parameters}. \textbf{a}, circular array of nano-antennas with varying width  from $1$ to $16$ in the range of $0 \leq \phi  \leq 2\pi $ \textbf{b}, Poincar\'e sphere plot of Stokes parameters of optical beams scattered from nano-antennas of (a) and (b) for LCP incidence beam. Color code corresponds to the nano-antenna number. \textbf{c,d}, spin and orbital angular momentum dependent beam profiles shown up as spin- and orbital-Hall effect. Scattered beam of RCP ($\sigma_{-}$) has the beam center of gravity shifted along $-x$-direction with asymmetric vortex beam profile for Gaussian incidence beam of LCP ($\sigma_{+}$) and the opposite shift and the vortex with sign of topological charge reversed in (d).
}
\end{figure*}

Consider a circular array of nano-antennas with different sizes from $1$ to $16$ in the range of $0 \le \phi \le 2\pi$, as shown in \textbf{Fig.~\ref{nonconstantPB} a}.
Scattering wave from the circular array acquires a phase factor originating from both  geometric phase $\gamma(\textbf{C})$ and dynamical phase $\Phi_D$,\cite{Berry45} where $\gamma(\textbf{C})$ is wavelength-independent while $\Phi_D$ has a frequency dispersion related to plasmonic resonances of each nano-antenna (see Supplementary Information).
When rotational symmetry is broken, azimuthal dynamical phase gradient $\nabla_\phi \Phi_{D}(\phi)$ is $\phi$-dependent, which results in $\phi$-dependent amplitude of scattered wave, generating additional vortex beams with $l \neq 2$.\cite{kotlyar2014asymmetricModes,kotlyar2014asymmetricBeams}
%
\begin{eqnarray}
E(r,\phi)_{-+} &=& BG_2(r)f(r,\phi) \exp\{i2\phi+ i\Phi_D^{-+}(\phi)\}\nonumber\\
&=& \sum_l^\infty \tilde{a}_l BG_l(r)e^{il\phi}\nonumber
\end{eqnarray}
where scattering wave is partial-wave expanded in term of vortex beams of topological charge $l$ with $BG_l (r) \equiv J_l(k_0r)e^{-{k_0}^2r^2}$, and  expansion coefficients $\tilde{a}_l={a}_le^{i\psi_l}$ are complex, and $f(r,\phi)=1$ when $\nabla_\phi \Phi_{D}^{-+}$ vanishes. A similar expression holds for $E(r,\phi)_{+-}$.

Owing to the difference in scattering amplitudes from nano-antennas with different sizes, polarization states of scattered waves form a spiral trajectory on Poincar\'e sphere as shown in \textbf{Fig.~\ref{nonconstantPB} b}. In other words, solid angles subtended by polarization states between two neighboring nano-antennas are not constant.
Rotational symmetry breaking introduced by nano-antennas with different sizes yields a spin-dependent non-constant azimuthal PB phase gradient, $\nabla_\phi \Phi_{PB}$, permitting spin-dependent deflections in SOC as shown in \textbf{Fig.~\ref{nonconstantPB} c}.

Analysis of PB phase in rotation symmetry-broken circular array of nano-antennas shows that
$\phi$-dependence of beam profiles $I_{-+}(r,\phi)$ and $I_{+-}(r,\phi)$ is determined by the total phase $\Phi(\phi)^{-+} = 2\phi + \Phi_D^{-+}(\phi)$ and $\Phi(\phi)^{+-} = -2\phi + \Phi_D^{+-}(\phi)$ acquired by cross polarization scattering beam, where non-constant azimuthal dynamical phase gradients $\nabla_\phi \Phi_{D}$ are responsible for $\phi$-dependence behavior of beam profiles. The occurrence of spin-dependent transverse shift, equivalent to non-vanishing $\Delta I \equiv I_{-+}-I_{+-}$, depends on the relation between $\Phi_D^{-+}(\phi)$ and $\Phi_D^{+-}(\phi)$ dictated by symmetry property of the circular array of nano-antennas with different sizes.

\section{Sample design and measurement}
\label{sec:sample}

\begin{figure*}[h]
\begin{center}
\includegraphics[width=15cm]{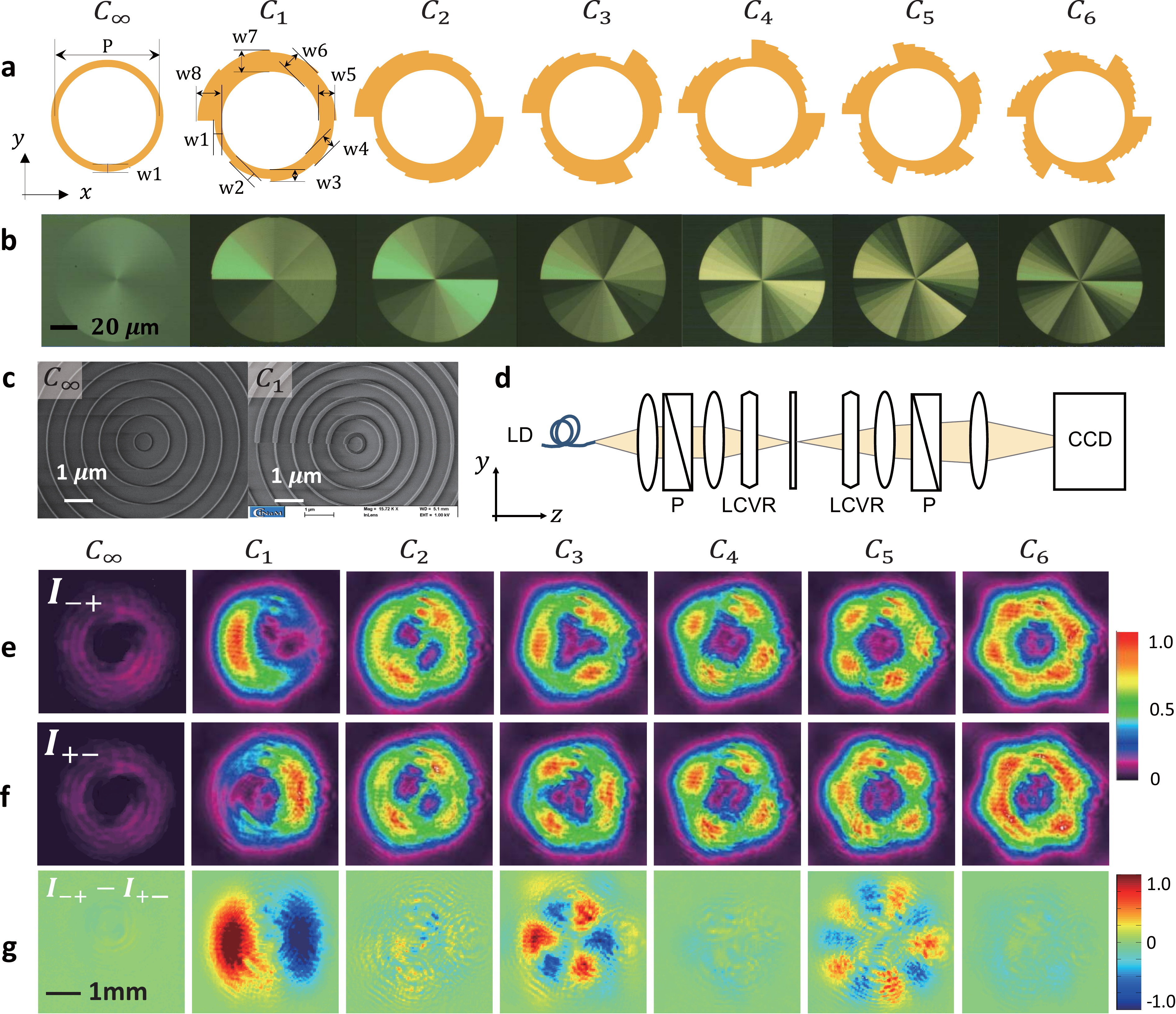}
\end{center}
\caption{\label{samplestructure}
\textbf{Cyclic group symmetric metasurfaces and extrinsic spin- and orbital-Hall effect}. \textbf{a}, Designed TA-CGSM with  $C_\infty$, $C_1$, $C_2$, $C_3$, $C_4$, $C_5$, and $C_6$ symmetry, respectively.  The periodicity is denoted by $p=600nm$, and widths of arcs are by $w_1=45nm$, $w_2=60nm$, $w_3=75nm$, $w_4=90nm$, $w_5=105nm$, $w_6=120nm$, $w_7= 135nm$, and $w_8=150nm$. \textbf{b}, Optical microscope images of fabricated TA-CGSMs belonging to  $C_\infty$, $C_1$, $C_2$, $C_3$, $C_4$, $C_5$, and $C_6$ symmetry, respectively. \textbf{c}, SEM images of fabricated TA-CGSMs belonging to $C_\infty$ and $C_1$ with scale bars of $  1  \mu m$. \textbf{d}, Experimental setup for spin- and orbital-Hall effect measurement. \textbf{e}, \textbf{f}, Far-field intensity distributions scattered from TA-CGSM of $C_{\infty}, C_{1}, C_{2}, C_{3}, C_{4}, C_{5},$ and $C_{6}$ for $I_{-+}$  and $I_{+-}$, respectively, measured with $\lambda=1310nm$ incidence beam.  \textbf{g}, Plot of the difference $I_{-+}-I_{+-}$ corresponding to extrinsic spin- and orbital-Hall effect.}
\end{figure*}

For a systematic study of relation between symmetry property of metasurface and spin-dependent transverse shift, tapered arc (TA) antennas are arranged in a circle to
form metasurfaces belonging to cyclic group $C_n$, called  tapered arc cyclic group symmetric metasurface (TA-CGSM), where in-plane inversion symmetry is determined by the order $n$ of cyclic group, as shown in \textbf{Fig.~\ref{samplestructure} a}.
%
%
Six different TA-CGSMs are fabricated, each TA-CSGM being composed of multiple TA antennas with  varying width from 45$nm$ to 150$nm$ organized in $8$ azimuthal segments of concentric rings repeated with 600$nm$ radial spacing.

We used 1$mm$ thick borosilicate glass round substrates with diameter of one inch (UQG Optics). First, the substrates were cleaned in sequential bathes of acetone and isopropanol, assisted by ultrasonicating. Then the substrates were rinsed in deionized water and dried by nitrogen flow followed by oxygen radical plasma treatment in a barrel reactor (Nanoplas, France) to activate the glass surface.
Second, an electron beam resist (PMMA, ARP 679) and a conductive polymer (SX AR PC 5000/90.2 from Allresist, Germany) layers were successively spin-coated. The thickness of the resulting layer was chosen to be about 70$nm$ and it was measured by a contact profilometer (DektakXT, Bruker, Germany) after e-beam exposition and resist development.
Third, we used an e-beam lithography (EBL) tool (Pioneer, Raith, Germany) for sample patterning. We used 20$kV$ acceleration voltage of the electron gun, and the beam current was 0.016$nA$ for the aperture of 7.5$\mu m$. The typical working distance was about 5$mm$, and the nominal exposition dose was chosen in the range from 100 to 180$\mu C/cm^2$.
Forth, two successive metal layers (Cr, 3$nm$ for seeding and Au, 27$nm$) were deposited by thermal evaporation (Auto 306, Edwards, UK). The thickness was monitored by quartz crystal microbalance during the deposition and measured by a contact profilometer after lift-off process. The lift-off was done in ethyl lactate solution, followed by rinsing and nitrogen drying.
Finally, the samples were observed by optical and scanning electron microscopy (SEM). For SEM observation, we used low acceleration voltage of 3$kV$ in order to decrease the sample charging because the glass substrate is insulating. (For more information, see Supplementary Information)

\textbf{Fig.~\ref{samplestructure} b} shows optical microscope images of TA-CGSMs possessing $C_\infty$, $C_1$, $C_2$, $C_3$, $C_4$, $C_5$, and $C_6$ symmetries, and SEM images of $C_\infty$ and $C_1$ TA-CGSMs are shown in \textbf{Fig.~\ref{samplestructure} c}.
%
%
Among experimental measurements in \textbf{Fig.~\ref{samplestructure} e, f}, $C_\infty$ TA-CGSM  exhibits degenerate $\phi$-independent intensity profiles for $I_{-+}$ and $I_{+-}$,
while $C_n (n\neq\infty)$ TA-CGSMs exhibits $\phi$-dependent intensity profiles of $I_{-+}$ and $I_{+-}$, owing to non-constant azimuthal PB phase gradient $\Phi_{PB}$ coming from plasmonic resonance of TAs.
Difference $\Delta I = I_{-+}-I_{+-}$ in \textbf{Fig.~\ref{samplestructure} g} corresponds to spin- and orbital-Hall shift, observed only in  $C_1$, $C_3,$ and $C_5$ TA-CGSMs (odd order $n$ of $C_n$), where degeneracy of geometric structure is lifted by in-plane inversion symmetry breaking.\cite{slobozhanyuk2016enhanced}
Vortex charge of cross-polarization scattering beam from $C_n (n\neq\infty)$ TA-CGSM is identified by an interference pattern between scattering vortex and incidence Gaussian beams,
and counter-clockwise and clockwise twisted fringes confirm topological charges of $l=2$ and $l=-2$ of $I_{-+}$ and $I_{+-}$, respectively. See \textbf{Fig.~S2} in Supplementary Information. FDTD calculation is in a perfect agreement with experimental results. See \textbf{Fig.~S1} in Supplementary Information.

\section{Partial wave decomposition}
\label{sec:partialwave}

\begin{figure*}[h]
\begin{center}
\includegraphics[width=15cm]{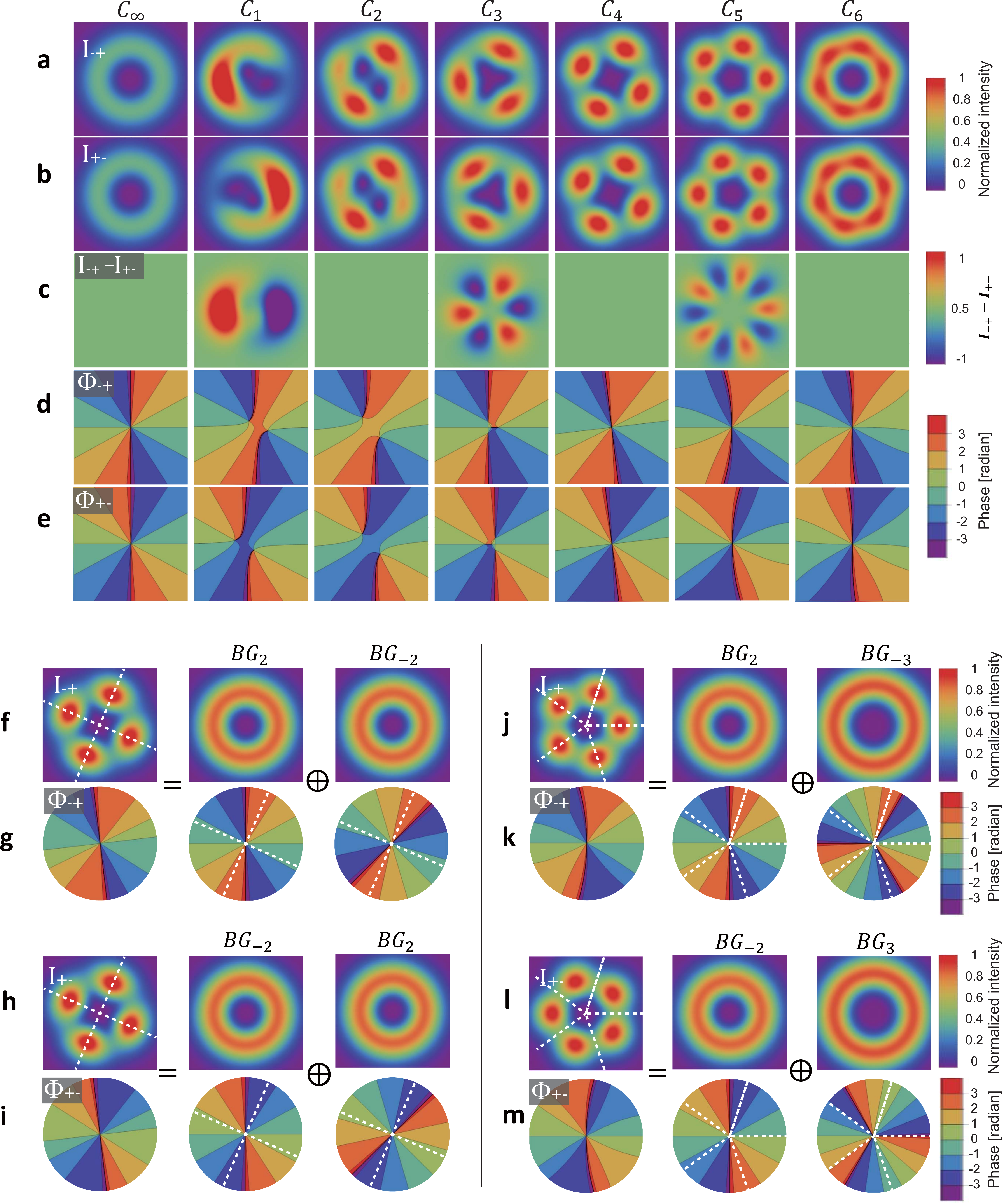}
\end{center}
\caption{\label{Results1cal}
\textbf{Analytically calculated extrinsic spin and orbital Hall effect for cyclic group symmetric metasurfaces}. Calculated far-field intensity (\textbf{a}, \textbf{b}). \textbf{c}, Calculated  $I_{-+}-I_{+-}$, and phase distributions (\textbf{d}, \textbf{e}) of spin flip component for $C_{\infty}, C_{1}, C_{2}, C_{3}, C_{4}, C_{5},$ and $C_{6}$ CGSM. \textbf{f-i} (\textbf{j-m}), Fourier decomposition of $E_{-+}$ and $E_{+-}$ of $C_{4}$ ($C_{5}$) CGSM.
}
\end{figure*}

Vortex beam with topological charge $l$ is described by Bessel-Gaussian beam,\cite{mendoza2015laguerre}
\begin{eqnarray}{{J}_{l}}(k_0r){{e}^{-{{{k_0}^2r}^{2}}}}\exp(i l \phi) = {BG}_{l}(r)\exp(i l \phi).\nonumber
\end{eqnarray}
In order to clarify how the addition of dynamical phase to geometric phase in SOC alters $\phi$-dependence of phases $\Phi^{-+}(\phi)$ and $\Phi^{+-}(\phi)$, experimentally measured intensity profiles in \textbf{Fig.~\ref{samplestructure} e, f} are fit as a linear superposition of vortex beams, as shown in \textbf{Fig.~\ref{Results1cal} a, b} along with phases in \textbf{Fig.~\ref{Results1cal} d, e}.
%
Spin-dependent vortex beams with asymmetric helical wave front are described by
\begin{eqnarray}
{{{E}}(r,\phi)}=\sum\limits_{l}{{{\tilde{a}}_{l}}}{{{BG}}_{l}(r)}{{{e}}^{i l\phi}}= A(r,\phi )e^{i{\Phi }_{\rm tot}},\nonumber
\end{eqnarray}
where $\tilde{a}_l$ is the $l$-th order complex coefficient of the Fourier expansion, i.e., $\tilde{a}_l={a}_l e^{i\psi_l}$, $BG_l (r) = J_l(k_{0}r)e^{-{k_{0}}^2r^2}$,  $A(r,\phi)=\left|E(r,\phi)\right|$ is the amplitude, ${{\Phi }_{\rm tot}}= \gamma (\mathbf{C})+{\Phi}_{D} = \tan^{-1}(\textit{Im}(E)/\textit{Re}(E))$ is total phase. It shows that the far fields are determined by the local topological features of interfering fields.
The fit values of partial-wave expansion coefficient $\tilde{a}_l$ are listed in \textbf{Table 1}.
%

%
We observe two important features.
First, phases of $E_{-+}(\phi)$ and $E_{+-}(\phi)$ are opposite, i.e., $\Phi^{-+}(\phi)= - \Phi^{+-}(\phi)$, which is from the opposite senses of rotation of polarization state trajectories on Poincar\'e sphere for $\{-+\}$ and $\{+-\}$. Second, intensity profile $I_{+-}(\phi)$  is the same as that of in-plane inversion symmetry operated $I_{-+}(\phi)$, i.e., $I_{+-}(\phi) = I_{-+}(\phi+\pi)$, as a result of $\Phi^{-+}(\phi)= - \Phi^{+-}(\phi)$. The two features immediately explain why spin- and orbital-Hall effect takes place only for $C_n$ with odd $n$. For even $n$, $\Delta I$ vanishes identically, since geometric structures possess in-plane inversion symmetry.

\begin{figure}[h]
\begin{center}
\includegraphics[width=15cm]{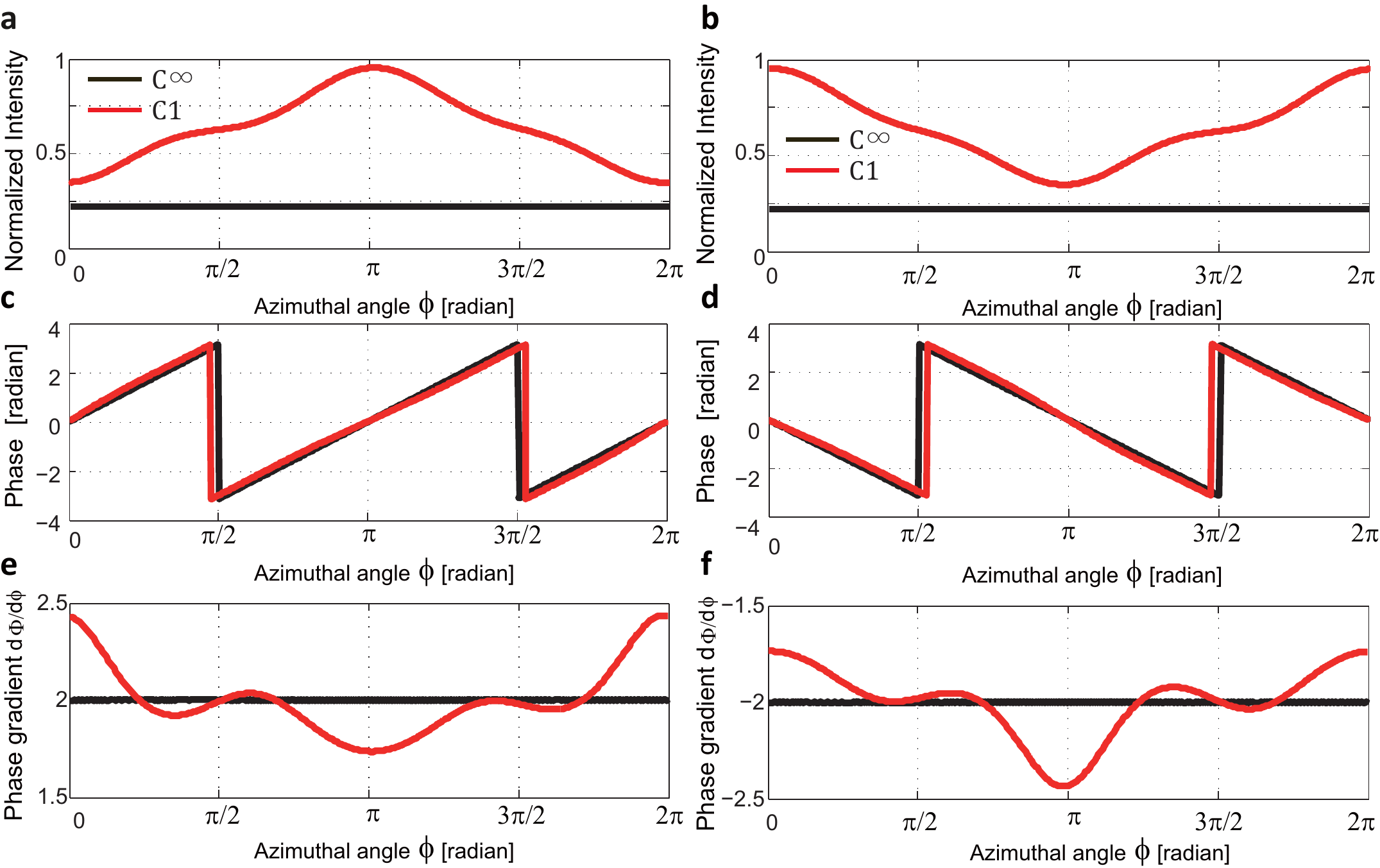}
\end{center}
\caption{\label{matlab}
\textbf{Analytically calculated intensity, phase, and phase gradient as a function of the azimuthal angle}.
\textbf{a,b}, Intensity $I_{ij}$, \textbf{c,d}, phase $\Phi_{ij}$, \textbf{e,f}, uniform phase gradient $d\Phi_{ij}/d\phi$ of $C_{\infty}$ (black) and nonuniform phase gradient of $C_{1}$ (red) for \{ij\} configuration, where \{ij\} are detection and incidence circular polarization helicity, \{-+\} for \textbf{a, c, e} and \{+-\} for \textbf{b, d, f}. The nonuniform phase gradient can be considered by introducing the anisotropy.}
\end{figure}

Now we further clarify the contributions of dynamical phase to spin-controlled spatial beam profiling in SOC.
First, we expect a correlation between intensity profile and $\phi$-dependence of phase. As shown in \textbf{Fig.~\ref{matlab}} for $C_1$ TA-CGSM,
$E_{-+}(\phi)$ and $E_{+-}(\phi)$  have a local maximum at the azimuthal angles $\phi$ where the absolute value  $|\nabla_\phi\Phi_D|$ is a local minimum. Helical wavefronts with azimuthal geometric phase gradient $\pm2$ are altered by introduction of non-constant azimuthal dynamical phase gradient $\nabla_\phi\Phi_D$, to build focused spots along azimuthal direction, similar to wavefront modification when a plane wave goes through a medium with spatially varying index.\cite{saleh1991fundamentals} Refer to Supplementary Information for  $C_n$ TA-CGSMs with $n$ other than $1$. The intensity profile possesses a feature of OAM superposition states, useful for quantum information systems.\cite{franke2008orbital}

Second, the phase $\psi_l$ of partial-wave expansion coefficient $\tilde{a}_l$ is related to dynamical phase $\Phi_D(\phi)$, determining interference behavior between partial-wave vortex beams having difference topological charge $l$s. $C_4$ and $C_5$ TA-CGSMs are closely examined in  the left panel (\textbf{Fig.~\ref{Results1cal} f, g, h, i}) and right panel (\textbf{Fig.~\ref{Results1cal} j, k, l, m}).
In $C_4$ TA-CGSM
\begin{eqnarray}
E_{-+} = BG_2 \exp\{i 2 \phi\}+ 0.24 BG_{-2}\exp\{-i2\phi\} \exp\{+i{3\pi\over 2}\}\nonumber
\end{eqnarray}
and
\begin{eqnarray}
E_{+-} = BG_{-2}\exp\{-i2\phi\} + 0.24 BG_{2}\exp\{i 2 \phi\} \exp\{-i{3\pi\over 2}\},\nonumber
\end{eqnarray}
where azimuthal angles for constructive and destructive interferences are the same for $E_{-+}$ and $E_{+-}$, resulting in $\Delta I = 0$.
In $C_5$ TA-CGSM
\begin{eqnarray}
E_{-+} = BG_2 \exp\{i2\phi\}+  BG_{-3}\exp\{-i3\phi\} \exp\{+i\pi\}\nonumber
\end{eqnarray}
and
\begin{eqnarray}
E_{+-} = BG_{-2}\exp\{-i2\phi\} + BG_{3}\exp\{i3\phi\} \exp\{-i\pi\},\nonumber
\end{eqnarray}
where azimuthal angles for constructive and destructive interferences differ by $0.2\pi$ for $E_{-+}$ and $E_{+-}$, and $\Delta I \neq 0$ leading to non-vanishing spin-dependent transverse shift.
Third, a frequency dispersion of dynamical phase due to the plasmonic resonance of TA renders intensity profiles to be dependent on the incidence beam wavelength, confirmed to be true.

\section{Wavelength dependence of extrinsic spin- and orbital-Hall effect}
\label{sec:wavelength}

\begin{figure*}[h]
\begin{center}
\includegraphics[width=15cm]{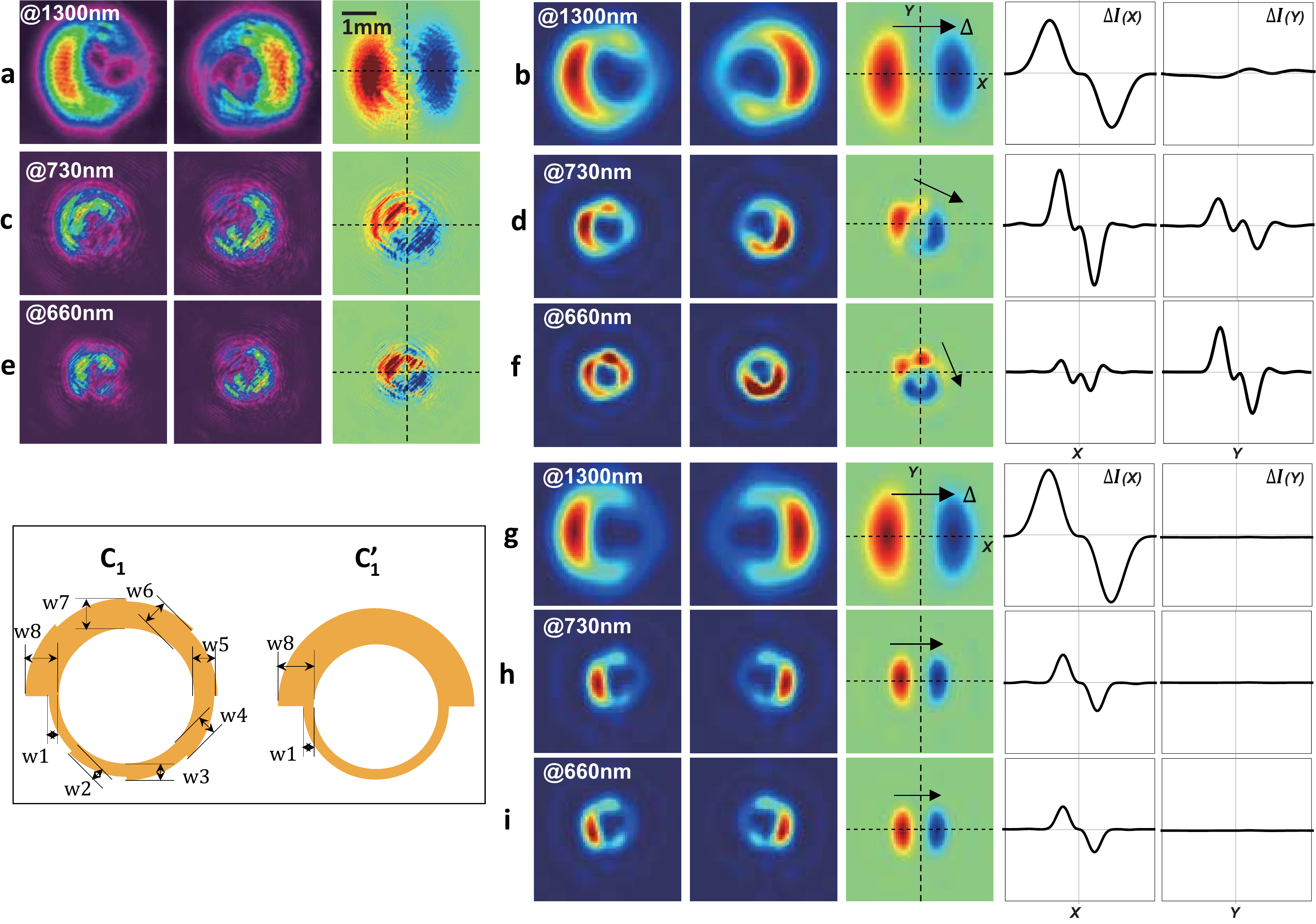}
\end{center}
\caption{\label{Results2}
\textbf{Wavelength dependence of extrinsic spin- and orbital- Hall effect}. Wavelength-dependence of spin- and orbital-Hall effect in $C_1$ metasurface is shown at 1300nm, 730nm,  and 550nm.  \textbf{a, c, e}, Experimental measurements and \textbf{b, d, f}, FDTD calculations for $C_1$ metasurface. \textbf{g,h,i}, FDTD calculations for ${C_1}'$ metasurface.
}
\end{figure*}

The role of dynamical phase can be identified by examining wavelength dependence of the spin- and orbital- Hall effect since plasmonic resonance of nano-antennas is wavelength-dependent.
We measured the spin- and orbital-dependence of far field intensity $C_1$ metasurface by employing lasers with different wavelengths as shown in \textbf{Fig. \ref{Results2}}.
Single-mode fiber pigtailed laser diodes with wavelengths $\lambda =1,310nm$, $\lambda =730nm$, and $\lambda =660nm$ are adopted as normal-incidence illumination sources.
Direction of the transverse shift goes through clock-wise rotation as the wavelength gets shorter. While geometric phase is wavelength-independent, dynamical phases originating from multiple TA antennas with varying widths give rise to wavelength- and $\phi$-dependent far-field intensity distributions.
%

On the other hand, ${C_1}'$ metasurface possessing $\sigma_{yz}$ symmetry, i.e., reflection symmetry with respect to $yz$-plane, allows for  splitting of vortex beam along $x$-direction only as shown in \textbf{Fig. \ref{Results2} g-i}, and  direction of the transverse shift is fixed along $x$-direction as the wavelength get shorter, exhibiting no wavelength-dependence. We note that in ${C_1}'$ only $y$-direction dynamical phase gradient gets involved owing to breaking of $y$-direction inversion-symmetry. Normalized horizontal $(X)$ and vertical $(Y)$ spin- and orbital-Hall effect profiles, $\Delta I (X)$ and $\Delta I (Y)$ along the dashed coordinate axes, are displayed in two right panel, where $\Delta I $=$I_{-+}(r,\phi)-I_{+-}(r,\phi)$.
\section{Conclusion}
\label{sec:conclusion}
In summary we demonstrated spin- and orbital-Hall effect in metasurfaces possessing cyclic point group symmetry $C_n$.
Lifting degeneracy in orbital angular momentum is achieved by selectively controlling the symmetry order $n$.
By identifying the role of dynamical phase in giving rise to spin-dependent transverse shift in SOC, spatial separation and modification of vortex beam profile are made possible.
This new approach provides significant advantages in applications such as vortex multiplexing in communication device and vortex beam analyzer as well as fundamental understating of interactions among angular momenta of light in metasurface.
\begin{table*}[t]
\centering
\label{my-label}
\begin{tabular}{|c|
>{\columncolor[HTML]{9B9B9B}}c ||c|c|c|c|c|c|c||c|c|c|c|c|c|c|}
\hline               & \multicolumn{1}{c||}{\cellcolor[HTML]{C0C0C0}\textbf{ }}
                     & \multicolumn{7}{c||}{\cellcolor[HTML]{C0C0C0}\textbf{$E_{-+}$}}                                                                                                                                                                                                                                                                                                    & \multicolumn{7}{c|}{\cellcolor[HTML]{C0C0C0}\textbf{$E_{+-}$}}                                                                                                                                                                                                                                                                                      \\ \hline
                     & \textbf{$l$}  & \cellcolor[HTML]{9B9B9B}\textbf{-4} & \cellcolor[HTML]{9B9B9B}\textbf{-3} & \cellcolor[HTML]{9B9B9B}\textbf{-2} & \cellcolor[HTML]{9B9B9B}\textbf{-1} & \cellcolor[HTML]{9B9B9B}\textbf{0} & \cellcolor[HTML]{9B9B9B}\textbf{1} & \cellcolor[HTML]{9B9B9B}\textbf{2} & \cellcolor[HTML]{9B9B9B}\textbf{-2} & \cellcolor[HTML]{9B9B9B}\textbf{-1} & \cellcolor[HTML]{9B9B9B}\textbf{0} & \cellcolor[HTML]{9B9B9B}\textbf{1} & \cellcolor[HTML]{9B9B9B}\textbf{2} & \cellcolor[HTML]{9B9B9B}\textbf{3} & \cellcolor[HTML]{9B9B9B}\textbf{4} \\ \hline\hline
                     & \textbf{${{{a}}_{l}}$} & 0                                   & 0                                   & 0                                   & 0.02                                & 0.01                               & 0.1                                & 1                                  & 1                                    & 0.1                                    & 0.01                                   & 0.02                                   & 0                                   & 0                                   & 0                                   \\ \cline{2-16}
\multirow{-2}{*}{$C_1$}
& \textbf{$\psi_{l}$}
& 0                                     & 0                                    & 0                                    & 0                                    & $\frac{\pi}{2}$                                  & $\pi$                                   & 0                                   & 0                                    & $-\pi$                                    & $-\frac{\pi}{2}$                                   & 0                                   & 0                                   & 0                                   & 0                                   \\ \hline
                     & \textbf{${{{a}}_{l}}$} & 0                                    & 0                                    & 0.12                                    & 0                                    & 0.02                                   & 0                                   & 1                                   & 1                                    & 0                                    & 0.02                                   & 0                                   & 0.12                                   & 0                                   & 0                                   \\ \cline{2-16}
\multirow{-2}{*}{$C_2$} &  \textbf{$\psi_{l}$}
& 0                                    & 0                                    & $\frac{3\pi}{2}$                                    & 0                                   & $\frac{\pi}{2}$                                   & 0                                   & 0                                   & 0                                    & 0                                   & $-\frac{\pi}{2}$                                   & 0                                   & $-\frac{3\pi}{2}$                                   & 0                                   & 0
\\ \hline
& \textbf{${{{a}}_{l}}$}
& 0                                    & 0                                    & 0                                    & 0.05                                    & 0                                   & 0                                   & 1                                   & 1                                    & 0                                    & 0                                   & 0.05                                   & 0                                   & 0                                   & 0                                   \\ \cline{2-16}
\multirow{-2}{*}{$C_3$} & \textbf{$\psi_{l}$}
& 0                                    & 0                                    & 0                                    & 0                                    & 0                                   & 0                                    & 0                                    & 0                                    & 0                                    & 0                                   & 0                                   & 0                                   & 0                                   & 0                                   \\ \hline
& \textbf{${{{a}}_{l}}$}
& 0                                    & 0                                    & 0.24                                    & 0                                    & 0                                   & 0                                   & 1                                   & 1                                    & 0                                    & 0                                   & 0                                   & 0.24                                   & 0
& 0                                   \\ \cline{2-16}
\multirow{-2}{*}{$C_4$} & \textbf{$\psi_{l}$}
& 0                                    & 0                                    & $\frac{3\pi}{2}$                                     & 0                                    & 0                                   & 0                                    & 0                                    & 0                                    & 0                                    & 0                                   & 0                                   & $-\frac{3\pi}{2}$                                   & 0                                   & 0                                   \\ \hline
&  \textbf{${{{a}}_{l}}$} & 0                                    & 1                                    & 0                                    & 0                                    & 0                                   & 0                                   & 1                                   & 1                                    & 0                                    & 0                                   & 0                                   & 0                                   & 1                                   & 0                                   \\ \cline{2-16}
\multirow{-2}{*}{$C_5$} & \textbf{$\psi_{l}$}
& 0                                     & $\pi$                                    & 0                                    & 0                                    & 0                                   & 0                                   & 0                                   & 0                                    & 0                                    & 0                                   & 0                                   & 0                                   & $-\pi$                                   & 0                                   \\ \hline
& \textbf{${{{a}}_{l}}$}
& 1.74                                     & 0                                    & 0                                    & 0                                    & 0                                   & 0                                   & 1                                   & 1                                   & 0                                    & 0                                   & 0                                   & 0                                   & 0                                   & 1.74                                   \\ \cline{2-16}
\multirow{-2}{*}{$C_6$} & \textbf{$\psi_{l}$}
& $\frac{\pi}{2}$                                    & 0                                    & 0                                    & 0                                    & 0                                   & 0                                   & 0                                    & 0                                    & 0                                    & 0                                   & 0                                   & 0                                   & 0                                   & $-\frac{\pi}{2}$                                   \\ \hline
\end{tabular}
\centering
\caption{Fit values of partial-wave expansion coefficient $\tilde{a}_l$, the $l$-th order complex coefficient}
\end{table*}

\clearpage
\section*{Acknowledgement}
JWW acknowledges the support from the Ministry of Science, ICT \& Future Planning, Korea (2015001948, 2014M3A6B3063706). Nanofabrication processes were performed in PLANETE cleanroom facility, CT PACA.

%

\clearpage

\clearpage

\section*{Supplementary Information}

\begin{enumerate}[label={\bf Section~{\arabic*}.}]

\item Finite difference time domain (FDTD) simulations

\item Interferogram with spherical wave

\item  Analytical study on wavelength dependent spin- and orbital-Hall effect

\item Calculated intensity, phase, and phase gradient for $C_{n}$ metasurfaces

\item Fourier decomposition in terms of Bessel-Gaussian beams

\item Sample fabrication

\item Calculation of Stokes parameters

\end{enumerate}

\clearpage
\setcounter{figure}{0}
\renewcommand{\thefigure}{\textbf{S\arabic{figure}}}
\def\theequation{S\arabic{equation}}

\begin{enumerate}[label={\bf Section~{\arabic*}.},leftmargin=*,align=left]
\item  \textbf{Finite difference time domain (FDTD) simulations}
\end{enumerate}

\begin{figure}[!h]
\begin{center}
\includegraphics[width=13cm]{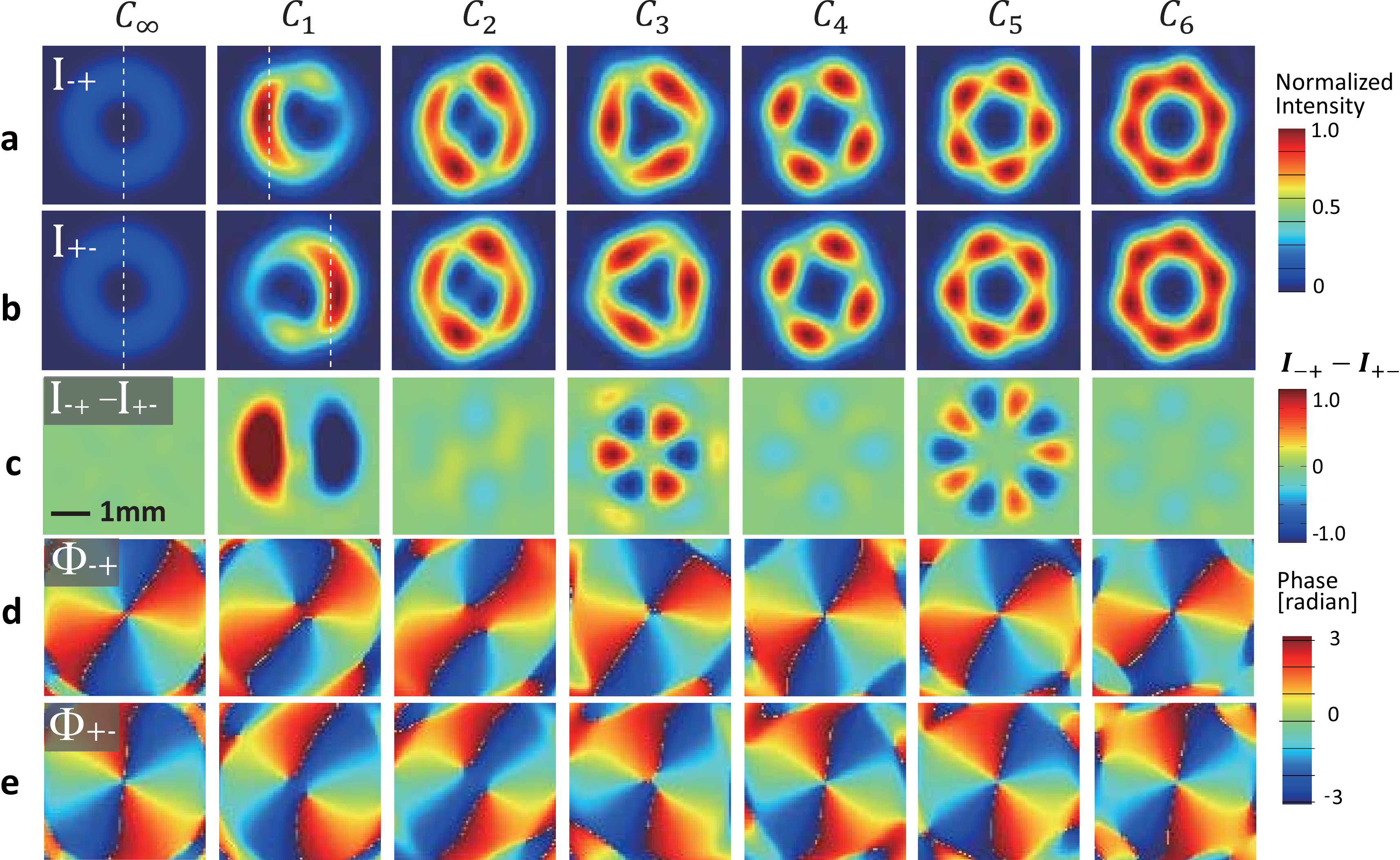}
\end{center}
\caption{\label{Results1-2}
\textbf{Extrinsic spin- and orbital-Hall effect for cyclic group symmetric metasurfaces}. FDTD Calculated far-field intensity (\textbf{a}, \textbf{b}). \textbf{c}, calculated  $I_{-+}-I_{+-}$, and phase distributions (\textbf{d}, \textbf{e}) of spin flip component for $C_{\infty}, C_{1}, C_{2}, C_{3}, C_{4}, C_{5}$, and $C_{6}$ metasurface. The white dashed lines indicate the centers of gravity calculated numerically from the intensity distributions.
}
\end{figure}

\noindent We used finite difference time domain (Lumerical FDTD) method to calculate the far-field intensity $(I_{-+} , I_{+-})$ and phase $({\Phi }_{-+}, {\Phi }_{+-})$ distributions of scattered field from $C_{n}$ metasurfaces. In \textbf{Fig.~\ref{Results1-2}}, the white dashed lines indicate the centers of gravity calculated numerically from the intensity distributions. $\sigma$- and $l$-dependent transverse shift of gravity corresponds to spin- and orbital-Hall effect of light.
\clearpage
\begin{enumerate}[label={\bf Section~{\arabic*}.},leftmargin=*,align=left]
\setcounter{enumi}{1}
\item \textbf{Interferogram with spherical wave}\\
\end{enumerate}

\begin{figure}[h]
\begin{center}
\includegraphics[width=14cm]{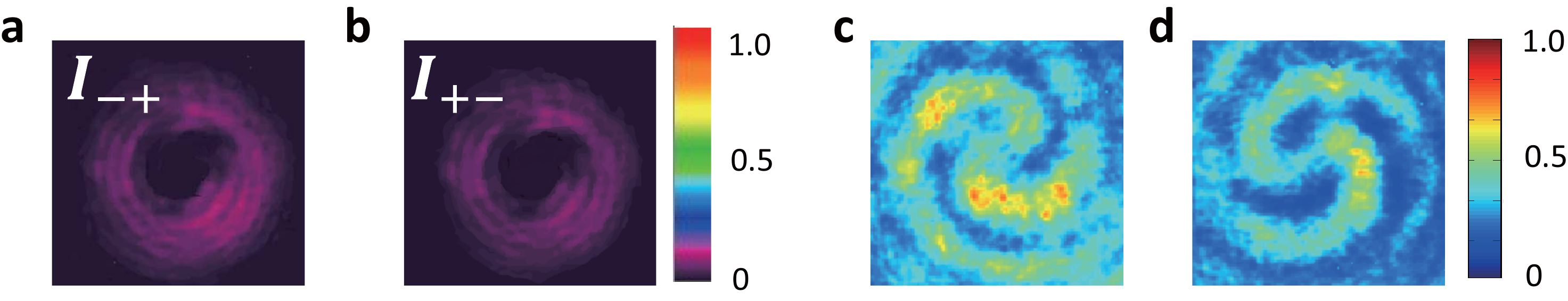}
\end{center}
\caption{\label{Results1-4}
\textbf{Measured intensity and interference patterns of $C_\infty$.} \textbf{a,b,} Measured transverse normalized intensity profile of Bessel-Gaussian beam with $l = \pm2$. \textbf{c,d,} Normalized interferogram with a Gaussian beam.
}
\end{figure}

Since the phase profile cannot be measured directly, the phase profile was obtained by taking an interferogram between helical beam and spherical wave. Counter-clockwise and clockwise twisted fringes indicate topological charge of $l=2$ and $l=-2$, respectively, as shown in \textbf{Fig.~\ref{Results1-4}}.


\clearpage
\begin{enumerate}[label={\bf Section~{\arabic*}.},leftmargin=*,align=left]
\setcounter{enumi}{2}
\item \textbf{Analytical study on wavelength dependent spin- and orbital-Hall effect }
\end{enumerate}

\begin{figure}[!h]
\begin{center}
\includegraphics[width=14cm]{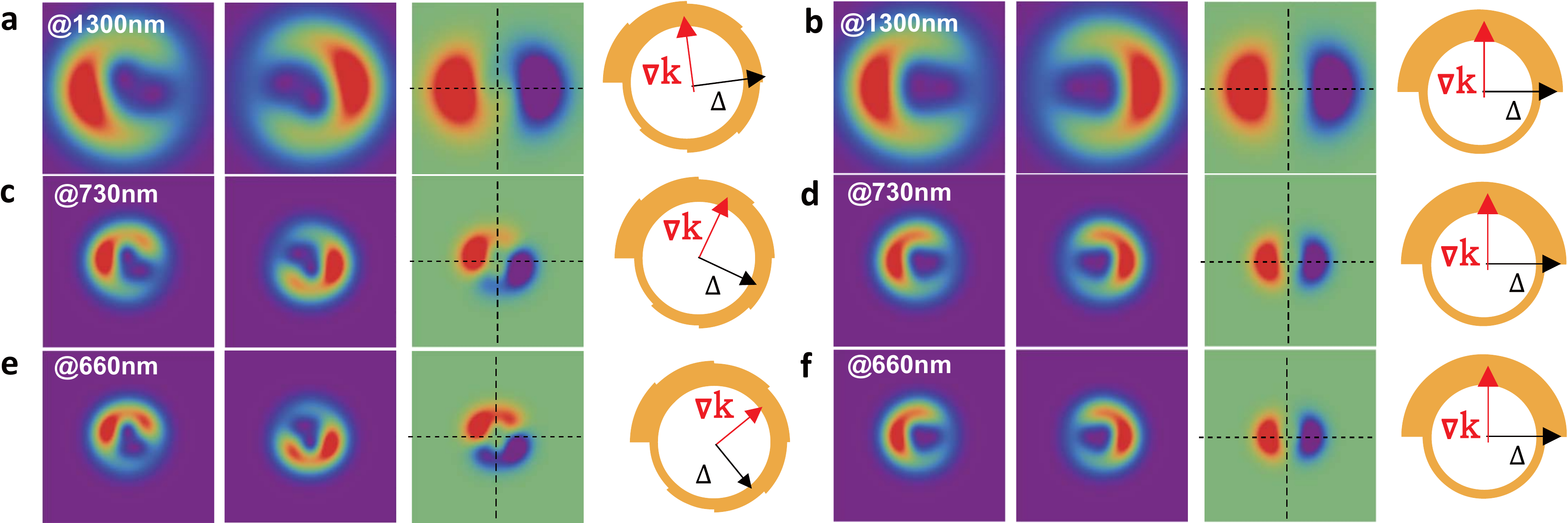}
\end{center}
\caption{\label{Results3}
\textbf{Wavelength dependent/independent spin- and orbital- Hall effect}. \textbf{a, c, e}, Calculated wavelength-dependent spin- and orbital-Hall effect in $C_1$ metasurface is shown at 1300nm, 730nm,  and 550nm. \textbf{b, d, f}, Calculated wavelength-independent spin- and orbital-Hall effect in ${C_1}'$ metasurface. 
}
\end{figure}

\textbf{Fig.\ref{Results3}} shows analytically calculated spin- and orbital-Hall effect for $C_1$ and $C_1'$ metasurfaces. In the case of the normal incidence of Gaussian beam with wavelength 1300nm, 730nm,  and 660nm, the transverse shift of cross polarization scattered light is described by black arrows, $\triangle$. The red arrows, $\nabla k$, indicate total tangential momentum gradient, i.e., spatial average of tangential component of Poynting vectors. From Lorentz force equation in momentum space, transverse shift $\triangle$ $\propto$  $\nabla k$ $\times k$. Therefore, it is made possible
to tailor wavelength-dependent transverse shift via wavefront engineering with plasmonic induced topological structure of dynamic phase. Partial wave expansion is shown in Table.

\begin{table}[h!]
\centering
\label{my-label}
\begin{tabular}{|c|
>{\columncolor[HTML]{9B9B9B}}c ||c|c|c|c|c||c|c|c|c|c|}
\hline               & \multicolumn{1}{c||}{\cellcolor[HTML]{C0C0C0}\textbf{1300nm}}
                     & \multicolumn{5}{c||}{\cellcolor[HTML]{C0C0C0}\textbf{$E_{-+}$}}                                                                                                                                                                                                                                                                                                    & \multicolumn{5}{c|}{\cellcolor[HTML]{C0C0C0}\textbf{$E_{+-}$}}                                                                                                                                                                                                                                                                                      \\ \hline
                     & \textbf{$l$}
                     & \cellcolor[HTML]{9B9B9B}\textbf{-2} & \cellcolor[HTML]{9B9B9B}\textbf{-1} & \cellcolor[HTML]{9B9B9B}\textbf{0} & \cellcolor[HTML]{9B9B9B}\textbf{1} & \cellcolor[HTML]{9B9B9B}\textbf{2} & \cellcolor[HTML]{9B9B9B}\textbf{-2} & \cellcolor[HTML]{9B9B9B}\textbf{-1} & \cellcolor[HTML]{9B9B9B}\textbf{0} & \cellcolor[HTML]{9B9B9B}\textbf{1} & \cellcolor[HTML]{9B9B9B}\textbf{2} \\ \hline\hline
                     & \textbf{${{{a}}_{l}}$}
                     & 0                                   & 0.02                                & 0.01                               & 0.1                                & 1                                  & 1                                    & 0.1                                    & 0.01                                   & 0.02                                   & 0             \\ \cline{2-12}
\multirow{-2}{*}{$C_1$}
& \textbf{$\psi_l$}
  & 0                                    & 0                                    & $\frac{\pi}{2}$                                  & $\pi$                                   & 0                                   & 0                                    & $-\pi$                                    & $\frac{-\pi}{2}$                                   & 0                                   & 0         \\ \hline
                     &\textbf{${{{a}}_{l}}$}
                     & 0                                   & 0.02                                    & 0                                   & 0.1                                   & 1                                   & 1                                    & 0.1                                   & 0                                   & 0.02                                   & 0              \\ \cline{2-12}
\multirow{-2}{*}{$C_1'$} &  \textbf{$\psi_l$}
& 0                                    & 0                                   & 0                                  & $\pi$                                   & 0                                   & 0                                    & $-\pi$                                   & 0                                   & 0
& 0 \\ \hline
\end{tabular}
\centering
\caption{Table for $l$-th order complex coefficient for $\lambda=1300nm$}
\end{table}
\begin{table}[h!]
\centering
\label{my-label}
\begin{tabular}{|c|
>{\columncolor[HTML]{9B9B9B}}c ||c|c|c|c|c||c|c|c|c|c|}
\hline               & \multicolumn{1}{c||}{\cellcolor[HTML]{C0C0C0}\textbf{730nm}}
                     & \multicolumn{5}{c||}{\cellcolor[HTML]{C0C0C0}\textbf{$E_{-+}$}}                                                                                                                                                                                                                                                                                                    & \multicolumn{5}{c|}{\cellcolor[HTML]{C0C0C0}\textbf{$E_{+-}$}}                                                                                                                                                                                                                                                                                      \\ \hline
                     & \textbf{$l$}
                     & \cellcolor[HTML]{9B9B9B}\textbf{-2} & \cellcolor[HTML]{9B9B9B}\textbf{-1} & \cellcolor[HTML]{9B9B9B}\textbf{0} & \cellcolor[HTML]{9B9B9B}\textbf{1} & \cellcolor[HTML]{9B9B9B}\textbf{2} & \cellcolor[HTML]{9B9B9B}\textbf{-2} & \cellcolor[HTML]{9B9B9B}\textbf{-1} & \cellcolor[HTML]{9B9B9B}\textbf{0} & \cellcolor[HTML]{9B9B9B}\textbf{1} & \cellcolor[HTML]{9B9B9B}\textbf{2} \\ \hline\hline
                     & \textbf{${{{a}}_{l}}$}
                     & 0                                   & 0.02                                & 0.01                               & 0.1                                & 1                                  & 1                                    & 0.1                                    & 0.01                                   & 0.02                                   & 0             \\ \cline{2-12}
\multirow{-2}{*}{$C_1$}
& \textbf{$\psi_l$}
  & 0                                    & $\frac{-\pi}{5}$                                    & $\frac{3\pi}{10}$                                  & $\frac{4\pi}{5}$                                    & 0                                   & 0                                    &  $\frac{-4\pi}{5}$                                    & $\frac{-3\pi}{10}$                                   & $\frac{\pi}{5}$                                   & 0         \\ \hline
                     &\textbf{${{{a}}_{l}}$}
                     & 0                                   & 0.02                                    & 0                                   & 0.1                                   & 1                                   & 1                                    & 0.1                                   & 0                                   & 0.02                                   & 0              \\ \cline{2-12}
\multirow{-2}{*}{$C_1'$} &  \textbf{$\psi_l$}
& 0                                    & $\frac{\pi}{5}$                                    & 0                                  & $\frac{4\pi}{5}$                                    & 0                                   & 0                                    & $\frac{-11\pi}{10}$                                    & 0                                   & $\frac{\pi}{5}$
& 0 \\ \hline
\end{tabular}
\centering
\caption{Table for $l$th-order complex coefficient for $\lambda=730nm$}
\centering
\label{my-label}
\begin{tabular}{|c|
>{\columncolor[HTML]{9B9B9B}}c ||c|c|c|c|c||c|c|c|c|c|}
\hline               & \multicolumn{1}{c||}{\cellcolor[HTML]{C0C0C0}\textbf{660nm}}
                     & \multicolumn{5}{c||}{\cellcolor[HTML]{C0C0C0}\textbf{$E_{-+}$}}                                                                                                                                                                                                                                                                                                    & \multicolumn{5}{c|}{\cellcolor[HTML]{C0C0C0}\textbf{$E_{+-}$}}                                                                                                                                                                                                                                                                                      \\ \hline
                     & \textbf{$l$}
                     & \cellcolor[HTML]{9B9B9B}\textbf{-2} & \cellcolor[HTML]{9B9B9B}\textbf{-1} & \cellcolor[HTML]{9B9B9B}\textbf{0} & \cellcolor[HTML]{9B9B9B}\textbf{1} & \cellcolor[HTML]{9B9B9B}\textbf{2} & \cellcolor[HTML]{9B9B9B}\textbf{-2} & \cellcolor[HTML]{9B9B9B}\textbf{-1} & \cellcolor[HTML]{9B9B9B}\textbf{0} & \cellcolor[HTML]{9B9B9B}\textbf{1} & \cellcolor[HTML]{9B9B9B}\textbf{2} \\ \hline\hline
                     & \textbf{${{{a}}_{l}}$}
                     & 0                                   & 0.02                                & 0.01                               & 0.1                                & 1                                  & 1                                    & 0.1                                    & 0.01                                   & 0.02                                   & 0             \\ \cline{2-12}
\multirow{-2}{*}{$C_1$}
& \textbf{$\psi_l$}
  & 0                                    & $\frac{-2\pi}{5}$                                    & $\frac{\pi}{10}$                                  & $\frac{3\pi}{5}$                                    & 0                                   & 0                                    &  $\frac{-3\pi}{5}$                                    & $\frac{-\pi}{10}$                                   & $\frac{2\pi}{5}$                                   & 0         \\ \hline
                     &\textbf{${{{a}}_{l}}$}
                     & 0                                   & 0.02                                    & 0                                   & 0.1                                   & 1                                   & 1                                    & 0.1                                   & 0                                   & 0.02                                   & 0              \\ \cline{2-12}
\multirow{-2}{*}{$C_1'$} &  \textbf{$\psi_l$}
& 0                                    & $\frac{\pi}{5}$                                    & 0                                  & $\frac{4\pi}{5}$                                    & 0                                   & 0                                    & $\frac{-11\pi}{10}$                                    & 0                                   & $\frac{\pi}{5}$
& 0 \\ \hline
\end{tabular}
\centering
\caption{Table for $l$th-order complex coefficient for $\lambda=660nm$}
\end{table}
\clearpage

\begin{enumerate}[label={\bf Section~{\arabic*}.},leftmargin=*,align=left]
\setcounter{enumi}{3}
\item  \textbf{Calculated intensity, phase, and phase gradient for $C_{n}$ metasurfaces}
\end{enumerate}
\begin{figure}[!h]
\begin{center}
\includegraphics[width=12cm]{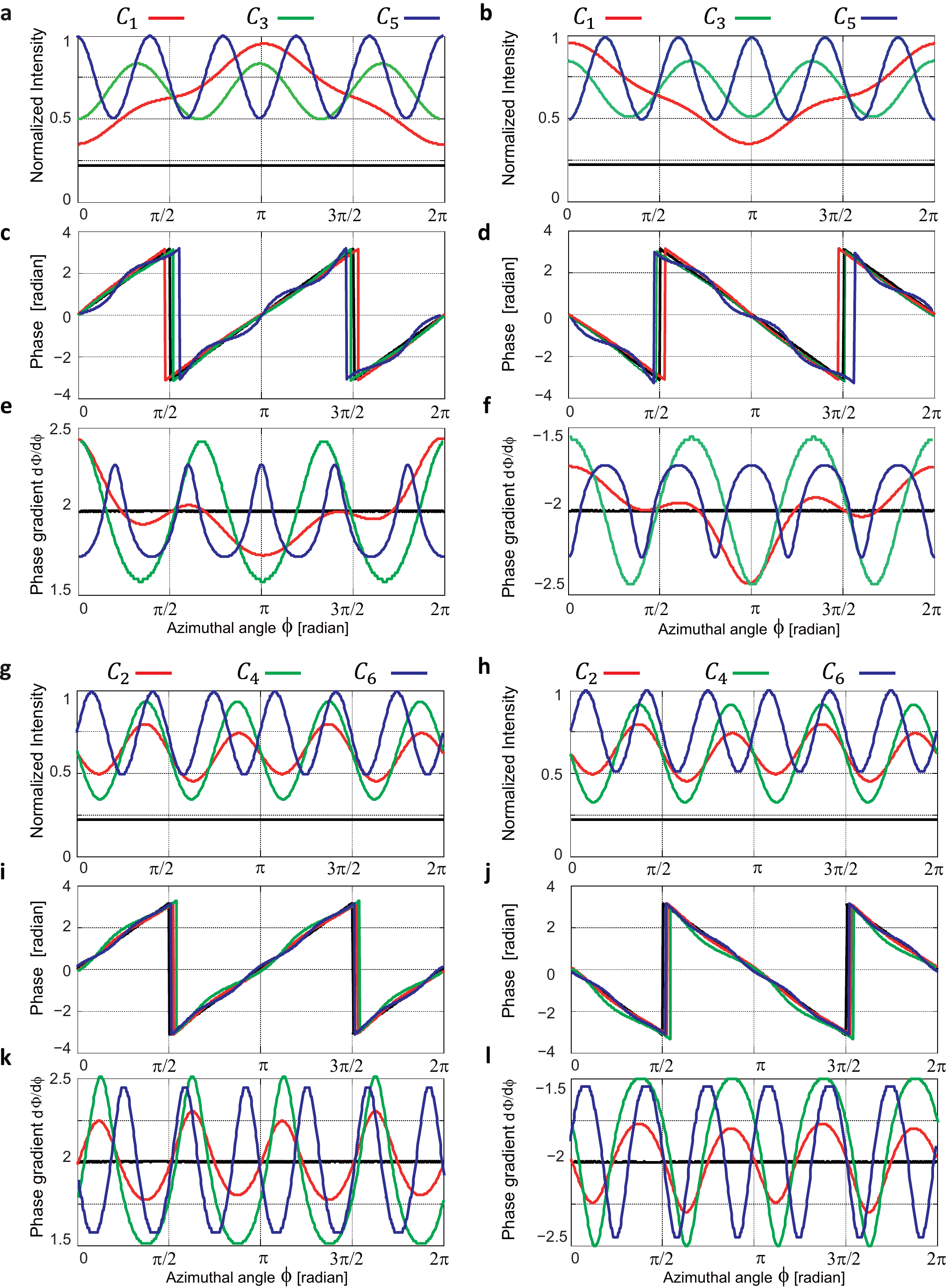}
\end{center}
\caption{\label{phase2}
\textbf{Analytically calculated intensity, phase, and phase gradient as a function of azimuthal angle}. \textbf{a,b}, Intensity, \textbf{c,d} phase, \textbf{e,f} phase gradient for $C_{1}$, $C_{3}$, $C_{5}$ TA-CGSM. \textbf{g,h}, Intensity, \textbf{i,j}, phase, \textbf{k,l} phase gradient for $C_{1}$, $C_{3}$, $C_{5}$ TA-CGSM for $\{-+\}$ and $\{+-\}$ configurations.
}
\end{figure}

\textbf{Fig.\ref{phase2}} shows analytically calculated intensity, phase, and phase gradient as a function of azimuthal angle for $C_{n}$ metasurfaces.
The net phase change in a closed path along the azimuthal direction is $4\pi$ and $-4\pi$ for \{-+\} and \{+-\} configurations as shown in  \textbf{Fig. \ref{phase2}} \textbf{ c, i} and \textbf{d, j}, respectively. Phase shows locally tailored features of non-constant phase gradient, introducing the displacement of beam center of gravity. According to linear momentum conservation, optical power flows along the phase change direction, and the azimuthal angle of a minimum of absolute value of phase corresponds to that of an optical power maximum shown in \textbf{Fig. \ref{phase2}} \textbf{a, g} and \textbf{b, h}. The results presented in \textbf{Fig. \ref{phase2}} clearly demonstrate that $I_{-+} (\phi) \neq I_{+-} (\phi)$ for $C_1, C_3,$ and $C_5$ metasurfaces corresponding to spin- and orbital-Hall effect of light. On the other hand, $I_{-+} (\phi) = I_{+-} (\phi)$ for $C_2, C_4,$ and $C_6$ metasurfaces possessing in-plane inversion symmetry.

\clearpage

\begin{enumerate}[label={\bf Section~{\arabic*}.},leftmargin=*,align=left]
\setcounter{enumi}{4}
\item  \textbf{Fourier decomposition in terms of Bessel-Gaussian beams}
\end{enumerate}
The spin-dependent vortex beams with asymmetric helical wave front are described by
\begin{equation}
{{{E}}(r,\phi)}=\sum\limits_{l}{{{\tilde{a}}_{l}}}{{{BG}}_{ l}}{{{e}}^{i l\phi}}\equiv A(r,\phi )e^{i{\Phi }_{tot}},
\end{equation}
where the $\tilde{a}_l$ is the $l$-th order complex coefficient of the Fourier expansion, i.e., $\tilde{a}_l={a}_l e^{i\psi_l}$, $BG_l (r) = J_l(k_{0}r)e^{-{k_{0}}^2r^2}$,  $A(r,\phi)=\left|E\right|$ amplitude , ${{\Phi }_{tot}}= \gamma (\mathbf{C})+{\Phi}_{D} = \tan^{-1}(\textit{Im}(E)/\textit{Re}(E))$ total phase. It shows that the far fields are determined by the local topological features of interfering fields.
Note that
\begin{eqnarray}
{\bf S}\equiv{BG_{2}(r)}{e^{i2\phi }}+{BG_{l}(r)}{e^{i(l\phi-\xi)}}= A(r,\phi )e^{i{\Phi }_{\rm tot}}\nonumber
\end{eqnarray}
where
\begin{eqnarray}
A(r,\phi)
=&&\sqrt{2B{{G}_{2}(r)}B{{G}_{l}(r)}\cos l(\xi-\phi )\cos (2\phi )-}\nonumber\\
&&\overline{2B{G_{2}(r)}B{{G}_{l}(r)}\sin l(\xi-\phi )\sin (2\phi )+B{{G}_{2}(r)}^{2}+B{{G}_{l}(r)}^{2}},\nonumber\\
\mbox{and}&&\nonumber\\%
{\Phi }_{\rm tot}=&&\tan^{-1}\left[\frac{ \textit{Im}({\bf S})}{ \textit{Re}({\bf S})}\right]=\tan^{-1}\left[ \frac{B{{G}_{2}(r)}\sin(2\phi)-B{{G}_{l}(r)}\sin l(\xi-\phi)}{B{{G}_{2}(r)}\cos(2\phi)+B{{G}_{l}(r)}\cos l(\xi-\phi)} \right].\nonumber
\end{eqnarray}


\begin{enumerate}[label={\bf Section~{\arabic*}.},leftmargin=*,align=left]
\setcounter{enumi}{5}
\item  \textbf{Sample fabrication}\\
\end{enumerate}
Plasmonic CGSMs were fabricated on 1$mm$ thick round borosilicate glass substrates with a diameter of 25$mm$ (WBO-215 from UQG Optics).
First, the substrates were cleaned in successive ultrasonic bathes of acetone and isopropyl alcohol (IPA, propan-2-ol) in order to remove eventual organic contaminants and dried under clean nitrogen flow. After chemical cleaning, the substrates were introduced into the barrel oxygen plasma reactor (Nanoplas, France) for 10 minutes at a temperature of 150$^{o}C$ in order to remove remaining solvent traces and to increase their wettability.
Second, 60-70$nm$ thick layer of e-beam resist (PMMA diluted in ethyl-lactate, AR-P 679 from All-Resist, Germany) was spin-coated onto the clean substrate surfaces at a rotation speed of 6000$rpm$. Then, the e-beam resist underwent a soft baking process (10 min at 170 $^{o}C$ on a hot plate), and the second, conductive resist was spin-coated on the e-beam resist (SX AR-PC 5000/90.1 from All-Resist) and baked for 2 min at 85 $^{o}C$. We used conductive resist in order to prevent charging of insulating glass surfaces during the electron-beam lithography (EBL) fabrication step. Pioneer system (from Raith, Germany) equipped with a field emission electron gun was used for EBL patterning. We used the e-beam acceleration voltage of 20$kV$, beam current of 0.016$nA$, and aperture of 7.5 $\mu m$. The working distance was about 5$mm$. Each sample contained a design of multiple CGSMs organized in a matrix with a pitch of 2$mm$ in order to prevent eventual interactions during optical measurements. We slightly varied the EBL exposure dose in order to finely tune the width of individual features constituting the CGSMs. In the matrix rows, the nominal exposure dose was increased from 130 to 180 $\mu C/cm^2$, and the order of different CGSMs increased from $C_1$ to $C_6$ in the matrix columns. PMMA is a positive resist, and the exposed resist areas are then easily dissolved in a corresponding solvent during the development step. After exposure, the conductive resist layer was removed with deionized water. The samples were then developed for 60 s in a commercial solution (AR 600-55 from All-Resist) containing a mixture of methyl isobutyl ketone (MIBK) and IPA. The development was stopped in pure IPA bath and the samples were dried under nitrogen flow.
Third, a 2-3$nm$ thick chromium seed layer and 27$nm$ thick optically active gold layer were successively evaporated under vacuum (Auto 306 tool from Edwards). The thickness of deposited metal was monitored in situ using a quartz crystal microbalance. A lift-off process in ethyl-lactate in an ultrasonic bath was used to remove the e-beam resist as well as the excess of gold and chromium in the sample areas which were not exposed to the electron beam during EBL. Finally, the samples were rinsed in deionized water and dried under nitrogen flow.

After fabrication of the samples, the total thickness of deposited metal layer (30$nm$) was confirmed by contact mechanical profilometer measurements (Dektak XT, Bruker, Germany). Then, the samples were characterized by optical microscopy as well as by Scanning Electron Microscopy (SEM) using the same Pioneer system as for EBL. All the CGSM were verified one by one by SEM observation in order to check for eventual (very rare) defects appeared during the lithography or lift-off steps. The sizes of individual nanofeatures were measured from SEM images taken at a low acceleration voltage (3$kV$) in order to decrease sample charging. Quick optical microscope observations were performed using Nikon optical microscope at Planete.


\begin{enumerate}[label={\bf Section~{\arabic*}.},leftmargin=*,align=left]
\setcounter{enumi}{6}
\item  \textbf{Calculation of Stokes parameters}
\end{enumerate}

To analyze vortex beam generation, we investigate Stokes parameters of space-variant nano-antennas as optical elements of TA-CGSMs on Poincar\'e sphere. On Poincar\'e sphere, any point in the equator, the north and south poles correspond to the linear polarization, right-circular and left-circular polarization state, respectively.
We calculated Stokes parameters of scattered light at wavelength of $1310 nm$ from each nano-antennas with finite difference time domain (FDTD) method. We defined the Stokes parameters as follows. $S_0$ is the total intensity, $S_1$ is the intensity of the horizontal linear component minus the intensity of the vertical linear component, $S_2$ is the intensity of the diagonal linear component minus the intensity of the antidiagonal linear component, and $S_3$ is the intensity of the right circular component minus the intensity of the left circular component. The normalized Stokes parameter $s_1=S_1/S_0, s_2=S_2/S_0$, and $s_3=S_3/S_0$ are plotted on the surface of Poincar\'e sphere.


\begin{thebibliography}{33}%
\makeatletter
\providecommand \@ifxundefined [1]{%
 \@ifx{#1\undefined}
}%
\providecommand \@ifnum [1]{%
 \ifnum #1\expandafter \@firstoftwo
 \else \expandafter \@secondoftwo
 \fi
}%
\providecommand \@ifx [1]{%
 \ifx #1\expandafter \@firstoftwo
 \else \expandafter \@secondoftwo
 \fi
}%
\providecommand \natexlab [1]{#1}%
\providecommand \enquote  [1]{``#1''}%
\providecommand \bibnamefont  [1]{#1}%
\providecommand \bibfnamefont [1]{#1}%
\providecommand \citenamefont [1]{#1}%
\providecommand \href@noop [0]{\@secondoftwo}%
\providecommand \href [0]{\begingroup \@sanitize@url \@href}%
\providecommand \@href[1]{\@@startlink{#1}\@@href}%
\providecommand \@@href[1]{\endgroup#1\@@endlink}%
\providecommand \@sanitize@url [0]{\catcode `\\12\catcode `\$12\catcode
  `\&12\catcode `\#12\catcode `\^12\catcode `\_12\catcode `\%12\relax}%
\providecommand \@@startlink[1]{}%
\providecommand \@@endlink[0]{}%
\providecommand \url  [0]{\begingroup\@sanitize@url \@url }%
\providecommand \@url [1]{\endgroup\@href {#1}{\urlprefix }}%
\providecommand \urlprefix  [0]{URL }%
\providecommand \Eprint [0]{\href }%
\providecommand \doibase [0]{http://dx.doi.org/}%
\providecommand \selectlanguage [0]{\@gobble}%
\providecommand \bibinfo  [0]{\@secondoftwo}%
\providecommand \bibfield  [0]{\@secondoftwo}%
\providecommand \translation [1]{[#1]}%
\providecommand \BibitemOpen [0]{}%
\providecommand \bibitemStop [0]{}%
\providecommand \bibitemNoStop [0]{.\EOS\space}%
\providecommand \EOS [0]{\spacefactor3000\relax}%
\providecommand \BibitemShut  [1]{\csname bibitem#1\endcsname}%
\let\auto@bib@innerbib\@empty
\bibitem [{\citenamefont {Allen}\ \emph {et~al.}(1992)\citenamefont {Allen},
  \citenamefont {Beijersbergen}, \citenamefont {Spreeuw},\ and\ \citenamefont
  {Woerdman}}]{allen1992orbital}%
  \BibitemOpen
  \bibfield  {author} {\bibinfo {author} {\bibfnamefont {L.}~\bibnamefont
  {Allen}}, \bibinfo {author} {\bibfnamefont {M.~W.}\ \bibnamefont
  {Beijersbergen}}, \bibinfo {author} {\bibfnamefont {R.}~\bibnamefont
  {Spreeuw}}, \ and\ \bibinfo {author} {\bibfnamefont {J.}~\bibnamefont
  {Woerdman}},\ }\href@noop {} {\bibfield  {journal} {\bibinfo  {journal}
  {Physical Review A}\ }\textbf {\bibinfo {volume} {45}},\ \bibinfo {pages}
  {8185} (\bibinfo {year} {1992})}\BibitemShut {NoStop}%
\bibitem [{\citenamefont {Bozinovic}\ \emph {et~al.}(2013)\citenamefont
  {Bozinovic}, \citenamefont {Yue}, \citenamefont {Ren}, \citenamefont {Tur},
  \citenamefont {Kristensen}, \citenamefont {Huang}, \citenamefont {Willner},\
  and\ \citenamefont {Ramachandran}}]{bozinovic2013terabit}%
  \BibitemOpen
  \bibfield  {author} {\bibinfo {author} {\bibfnamefont {N.}~\bibnamefont
  {Bozinovic}}, \bibinfo {author} {\bibfnamefont {Y.}~\bibnamefont {Yue}},
  \bibinfo {author} {\bibfnamefont {Y.}~\bibnamefont {Ren}}, \bibinfo {author}
  {\bibfnamefont {M.}~\bibnamefont {Tur}}, \bibinfo {author} {\bibfnamefont
  {P.}~\bibnamefont {Kristensen}}, \bibinfo {author} {\bibfnamefont
  {H.}~\bibnamefont {Huang}}, \bibinfo {author} {\bibfnamefont {A.~E.}\
  \bibnamefont {Willner}}, \ and\ \bibinfo {author} {\bibfnamefont
  {S.}~\bibnamefont {Ramachandran}},\ }\href@noop {} {\bibfield  {journal}
  {\bibinfo  {journal} {Science}\ }\textbf {\bibinfo {volume} {340}},\ \bibinfo
  {pages} {1545} (\bibinfo {year} {2013})}\BibitemShut {NoStop}%
\bibitem [{\citenamefont {Ren}\ \emph {et~al.}(2016)\citenamefont {Ren},
  \citenamefont {Li}, \citenamefont {Zhang},\ and\ \citenamefont
  {Gu}}]{ren2016chip}%
  \BibitemOpen
  \bibfield  {author} {\bibinfo {author} {\bibfnamefont {H.}~\bibnamefont
  {Ren}}, \bibinfo {author} {\bibfnamefont {X.}~\bibnamefont {Li}}, \bibinfo
  {author} {\bibfnamefont {Q.}~\bibnamefont {Zhang}}, \ and\ \bibinfo {author}
  {\bibfnamefont {M.}~\bibnamefont {Gu}},\ }\href@noop {} {\bibfield  {journal}
  {\bibinfo  {journal} {Science}\ }\textbf {\bibinfo {volume} {352}},\ \bibinfo
  {pages} {805} (\bibinfo {year} {2016})}\BibitemShut {NoStop}%
\bibitem [{\citenamefont {Yang}\ \emph {et~al.}(2016)\citenamefont {Yang},
  \citenamefont {Pu}, \citenamefont {Li}, \citenamefont {Ma}, \citenamefont
  {Luo}, \citenamefont {Gao},\ and\ \citenamefont {Luo}}]{yang2016wavelength}%
  \BibitemOpen
  \bibfield  {author} {\bibinfo {author} {\bibfnamefont {K.}~\bibnamefont
  {Yang}}, \bibinfo {author} {\bibfnamefont {M.}~\bibnamefont {Pu}}, \bibinfo
  {author} {\bibfnamefont {X.}~\bibnamefont {Li}}, \bibinfo {author}
  {\bibfnamefont {X.}~\bibnamefont {Ma}}, \bibinfo {author} {\bibfnamefont
  {J.}~\bibnamefont {Luo}}, \bibinfo {author} {\bibfnamefont {H.}~\bibnamefont
  {Gao}}, \ and\ \bibinfo {author} {\bibfnamefont {X.}~\bibnamefont {Luo}},\
  }\href@noop {} {\bibfield  {journal} {\bibinfo  {journal} {Nanoscale}\ }
  (\bibinfo {year} {2016})}\BibitemShut {NoStop}%
\bibitem [{\citenamefont {Bliokh}\ \emph {et~al.}(2015)\citenamefont {Bliokh},
  \citenamefont {Rodr{\'\i}guez-Fortu{\~n}o}, \citenamefont {Nori},\ and\
  \citenamefont {Zayats}}]{bliokh2015spin}%
  \BibitemOpen
  \bibfield  {author} {\bibinfo {author} {\bibfnamefont {K.}~\bibnamefont
  {Bliokh}}, \bibinfo {author} {\bibfnamefont {F.}~\bibnamefont
  {Rodr{\'\i}guez-Fortu{\~n}o}}, \bibinfo {author} {\bibfnamefont
  {F.}~\bibnamefont {Nori}}, \ and\ \bibinfo {author} {\bibfnamefont {A.~V.}\
  \bibnamefont {Zayats}},\ }\href@noop {} {\bibfield  {journal} {\bibinfo
  {journal} {Nature Photonics}\ }\textbf {\bibinfo {volume} {9}},\ \bibinfo
  {pages} {796} (\bibinfo {year} {2015})}\BibitemShut {NoStop}%
\bibitem [{\citenamefont {Allen}\ \emph {et~al.}(1999)\citenamefont {Allen},
  \citenamefont {Padgett},\ and\ \citenamefont {Babiker}}]{allen1999iv}%
  \BibitemOpen
  \bibfield  {author} {\bibinfo {author} {\bibfnamefont {L.}~\bibnamefont
  {Allen}}, \bibinfo {author} {\bibfnamefont {M.}~\bibnamefont {Padgett}}, \
  and\ \bibinfo {author} {\bibfnamefont {M.}~\bibnamefont {Babiker}},\
  }\href@noop {} {\bibfield  {journal} {\bibinfo  {journal} {Progress in
  optics}\ }\textbf {\bibinfo {volume} {39}},\ \bibinfo {pages} {291} (\bibinfo
  {year} {1999})}\BibitemShut {NoStop}%
\bibitem [{\citenamefont {Padgett}\ \emph {et~al.}(2004)\citenamefont
  {Padgett}, \citenamefont {Courtial},\ and\ \citenamefont
  {Allen}}]{padgett2004light}%
  \BibitemOpen
  \bibfield  {author} {\bibinfo {author} {\bibfnamefont {M.}~\bibnamefont
  {Padgett}}, \bibinfo {author} {\bibfnamefont {J.}~\bibnamefont {Courtial}}, \
  and\ \bibinfo {author} {\bibfnamefont {L.}~\bibnamefont {Allen}},\
  }\href@noop {} {\bibfield  {journal} {\bibinfo  {journal} {Physics Today}\
  }\textbf {\bibinfo {volume} {57}},\ \bibinfo {pages} {35} (\bibinfo {year}
  {2004})}\BibitemShut {NoStop}%
\bibitem [{\citenamefont {Cai}\ \emph {et~al.}(2012)\citenamefont {Cai},
  \citenamefont {Wang}, \citenamefont {Strain}, \citenamefont {Johnson-Morris},
  \citenamefont {Zhu}, \citenamefont {Sorel}, \citenamefont {O’Brien},
  \citenamefont {Thompson},\ and\ \citenamefont {Yu}}]{cai2012integrated}%
  \BibitemOpen
  \bibfield  {author} {\bibinfo {author} {\bibfnamefont {X.}~\bibnamefont
  {Cai}}, \bibinfo {author} {\bibfnamefont {J.}~\bibnamefont {Wang}}, \bibinfo
  {author} {\bibfnamefont {M.~J.}\ \bibnamefont {Strain}}, \bibinfo {author}
  {\bibfnamefont {B.}~\bibnamefont {Johnson-Morris}}, \bibinfo {author}
  {\bibfnamefont {J.}~\bibnamefont {Zhu}}, \bibinfo {author} {\bibfnamefont
  {M.}~\bibnamefont {Sorel}}, \bibinfo {author} {\bibfnamefont {J.~L.}\
  \bibnamefont {O’Brien}}, \bibinfo {author} {\bibfnamefont {M.~G.}\
  \bibnamefont {Thompson}}, \ and\ \bibinfo {author} {\bibfnamefont
  {S.}~\bibnamefont {Yu}},\ }\href@noop {} {\bibfield  {journal} {\bibinfo
  {journal} {Science}\ }\textbf {\bibinfo {volume} {338}},\ \bibinfo {pages}
  {363} (\bibinfo {year} {2012})}\BibitemShut {NoStop}%
\bibitem [{\citenamefont {Li}\ \emph {et~al.}(2015)\citenamefont {Li},
  \citenamefont {Phillips}, \citenamefont {Wang}, \citenamefont {Ho},
  \citenamefont {Chen}, \citenamefont {Zhou}, \citenamefont {Zhu},
  \citenamefont {Yu},\ and\ \citenamefont {Cai}}]{li2015orbital}%
  \BibitemOpen
  \bibfield  {author} {\bibinfo {author} {\bibfnamefont {H.}~\bibnamefont
  {Li}}, \bibinfo {author} {\bibfnamefont {D.~B.}\ \bibnamefont {Phillips}},
  \bibinfo {author} {\bibfnamefont {X.}~\bibnamefont {Wang}}, \bibinfo {author}
  {\bibfnamefont {Y.-L.~D.}\ \bibnamefont {Ho}}, \bibinfo {author}
  {\bibfnamefont {L.}~\bibnamefont {Chen}}, \bibinfo {author} {\bibfnamefont
  {X.}~\bibnamefont {Zhou}}, \bibinfo {author} {\bibfnamefont {J.}~\bibnamefont
  {Zhu}}, \bibinfo {author} {\bibfnamefont {S.}~\bibnamefont {Yu}}, \ and\
  \bibinfo {author} {\bibfnamefont {X.}~\bibnamefont {Cai}},\ }\href@noop {}
  {\bibfield  {journal} {\bibinfo  {journal} {Optica}\ }\textbf {\bibinfo
  {volume} {2}},\ \bibinfo {pages} {547} (\bibinfo {year} {2015})}\BibitemShut
  {NoStop}%
\bibitem [{\citenamefont {Biener}\ \emph {et~al.}(2002)\citenamefont {Biener},
  \citenamefont {Niv}, \citenamefont {Kleiner},\ and\ \citenamefont
  {Hasman}}]{biener2002formation}%
  \BibitemOpen
  \bibfield  {author} {\bibinfo {author} {\bibfnamefont {G.}~\bibnamefont
  {Biener}}, \bibinfo {author} {\bibfnamefont {A.}~\bibnamefont {Niv}},
  \bibinfo {author} {\bibfnamefont {V.}~\bibnamefont {Kleiner}}, \ and\
  \bibinfo {author} {\bibfnamefont {E.}~\bibnamefont {Hasman}},\ }\href@noop {}
  {\bibfield  {journal} {\bibinfo  {journal} {Optics letters}\ }\textbf
  {\bibinfo {volume} {27}},\ \bibinfo {pages} {1875} (\bibinfo {year}
  {2002})}\BibitemShut {NoStop}%
\bibitem [{\citenamefont {Marrucci}\ \emph {et~al.}(2006)\citenamefont
  {Marrucci}, \citenamefont {Manzo},\ and\ \citenamefont
  {Paparo}}]{marrucci2006optical}%
  \BibitemOpen
  \bibfield  {author} {\bibinfo {author} {\bibfnamefont {L.}~\bibnamefont
  {Marrucci}}, \bibinfo {author} {\bibfnamefont {C.}~\bibnamefont {Manzo}}, \
  and\ \bibinfo {author} {\bibfnamefont {D.}~\bibnamefont {Paparo}},\
  }\href@noop {} {\bibfield  {journal} {\bibinfo  {journal} {Physical review
  letters}\ }\textbf {\bibinfo {volume} {96}},\ \bibinfo {pages} {163905}
  (\bibinfo {year} {2006})}\BibitemShut {NoStop}%
\bibitem [{\citenamefont {Devlin}\ \emph {et~al.}(2016)\citenamefont {Devlin},
  \citenamefont {Ambrosio}, \citenamefont {Wintz}, \citenamefont {Oscurato},
  \citenamefont {Zhu}, \citenamefont {Khorasaninejad}, \citenamefont {Oh},
  \citenamefont {Maddalena},\ and\ \citenamefont {Capasso}}]{devlin2016spin}%
  \BibitemOpen
  \bibfield  {author} {\bibinfo {author} {\bibfnamefont {R.~C.}\ \bibnamefont
  {Devlin}}, \bibinfo {author} {\bibfnamefont {A.}~\bibnamefont {Ambrosio}},
  \bibinfo {author} {\bibfnamefont {D.}~\bibnamefont {Wintz}}, \bibinfo
  {author} {\bibfnamefont {S.~L.}\ \bibnamefont {Oscurato}}, \bibinfo {author}
  {\bibfnamefont {A.~Y.}\ \bibnamefont {Zhu}}, \bibinfo {author} {\bibfnamefont
  {M.}~\bibnamefont {Khorasaninejad}}, \bibinfo {author} {\bibfnamefont
  {J.}~\bibnamefont {Oh}}, \bibinfo {author} {\bibfnamefont {P.}~\bibnamefont
  {Maddalena}}, \ and\ \bibinfo {author} {\bibfnamefont {F.}~\bibnamefont
  {Capasso}},\ }\href@noop {} {\bibfield  {journal} {\bibinfo  {journal} {arXiv
  preprint arXiv:1605.03899}\ } (\bibinfo {year} {2016})}\BibitemShut {NoStop}%
\bibitem [{\citenamefont {Bliokh}\ \emph {et~al.}(2010)\citenamefont {Bliokh},
  \citenamefont {Alonso}, \citenamefont {Ostrovskaya},\ and\ \citenamefont
  {Aiello}}]{bliokh2010angular}%
  \BibitemOpen
  \bibfield  {author} {\bibinfo {author} {\bibfnamefont {K.~Y.}\ \bibnamefont
  {Bliokh}}, \bibinfo {author} {\bibfnamefont {M.~A.}\ \bibnamefont {Alonso}},
  \bibinfo {author} {\bibfnamefont {E.~A.}\ \bibnamefont {Ostrovskaya}}, \ and\
  \bibinfo {author} {\bibfnamefont {A.}~\bibnamefont {Aiello}},\ }\href@noop {}
  {\bibfield  {journal} {\bibinfo  {journal} {Physical Review A}\ }\textbf
  {\bibinfo {volume} {82}},\ \bibinfo {pages} {063825} (\bibinfo {year}
  {2010})}\BibitemShut {NoStop}%
\bibitem [{\citenamefont {Shitrit}\ \emph
  {et~al.}(2013{\natexlab{a}})\citenamefont {Shitrit}, \citenamefont
  {Yulevich}, \citenamefont {Maguid}, \citenamefont {Ozeri}, \citenamefont
  {Veksler}, \citenamefont {Kleiner},\ and\ \citenamefont
  {Hasman}}]{shitrit2013spin}%
  \BibitemOpen
  \bibfield  {author} {\bibinfo {author} {\bibfnamefont {N.}~\bibnamefont
  {Shitrit}}, \bibinfo {author} {\bibfnamefont {I.}~\bibnamefont {Yulevich}},
  \bibinfo {author} {\bibfnamefont {E.}~\bibnamefont {Maguid}}, \bibinfo
  {author} {\bibfnamefont {D.}~\bibnamefont {Ozeri}}, \bibinfo {author}
  {\bibfnamefont {D.}~\bibnamefont {Veksler}}, \bibinfo {author} {\bibfnamefont
  {V.}~\bibnamefont {Kleiner}}, \ and\ \bibinfo {author} {\bibfnamefont
  {E.}~\bibnamefont {Hasman}},\ }\href@noop {} {\bibfield  {journal} {\bibinfo
  {journal} {Science}\ }\textbf {\bibinfo {volume} {340}},\ \bibinfo {pages}
  {724} (\bibinfo {year} {2013}{\natexlab{a}})}\BibitemShut {NoStop}%
\bibitem [{\citenamefont {Cardano}\ and\ \citenamefont
  {Marrucci}(2015)}]{cardano2015spin}%
  \BibitemOpen
  \bibfield  {author} {\bibinfo {author} {\bibfnamefont {F.}~\bibnamefont
  {Cardano}}\ and\ \bibinfo {author} {\bibfnamefont {L.}~\bibnamefont
  {Marrucci}},\ }\href@noop {} {\bibfield  {journal} {\bibinfo  {journal}
  {Nature Photonics}\ }\textbf {\bibinfo {volume} {9}},\ \bibinfo {pages} {776}
  (\bibinfo {year} {2015})}\BibitemShut {NoStop}%
\bibitem [{\citenamefont {Berry}(1989)}]{Berry-BookChapter}%
  \BibitemOpen
  \bibfield  {author} {\bibinfo {author} {\bibfnamefont {M.}~\bibnamefont
  {Berry}},\ }in\ \href@noop {} {\emph {\bibinfo {booktitle} {Geometric phases
  in physics}}},\ \bibinfo {editor} {edited by\ \bibinfo {editor}
  {\bibfnamefont {D.~L.}\ \bibnamefont {Andrews}}\ and\ \bibinfo {editor}
  {\bibfnamefont {M.}~\bibnamefont {Babiker}}}\ (\bibinfo  {publisher} {World
  Scientific Singapore},\ \bibinfo {year} {1989})\BibitemShut {NoStop}%
\bibitem [{\citenamefont {Onoda}\ \emph {et~al.}(2004)\citenamefont {Onoda},
  \citenamefont {Murakami},\ and\ \citenamefont {Nagaosa}}]{onoda2004hall}%
  \BibitemOpen
  \bibfield  {author} {\bibinfo {author} {\bibfnamefont {M.}~\bibnamefont
  {Onoda}}, \bibinfo {author} {\bibfnamefont {S.}~\bibnamefont {Murakami}}, \
  and\ \bibinfo {author} {\bibfnamefont {N.}~\bibnamefont {Nagaosa}},\
  }\href@noop {} {\bibfield  {journal} {\bibinfo  {journal} {Physical review
  letters}\ }\textbf {\bibinfo {volume} {93}},\ \bibinfo {pages} {083901}
  (\bibinfo {year} {2004})}\BibitemShut {NoStop}%
\bibitem [{\citenamefont {Bliokh}\ \emph {et~al.}(2008)\citenamefont {Bliokh},
  \citenamefont {Niv}, \citenamefont {Kleiner},\ and\ \citenamefont
  {Hasman}}]{bliokh2008geometrodynamics}%
  \BibitemOpen
  \bibfield  {author} {\bibinfo {author} {\bibfnamefont {K.~Y.}\ \bibnamefont
  {Bliokh}}, \bibinfo {author} {\bibfnamefont {A.}~\bibnamefont {Niv}},
  \bibinfo {author} {\bibfnamefont {V.}~\bibnamefont {Kleiner}}, \ and\
  \bibinfo {author} {\bibfnamefont {E.}~\bibnamefont {Hasman}},\ }\href@noop {}
  {\bibfield  {journal} {\bibinfo  {journal} {Nature Photonics}\ }\textbf
  {\bibinfo {volume} {2}},\ \bibinfo {pages} {748} (\bibinfo {year}
  {2008})}\BibitemShut {NoStop}%
\bibitem [{\citenamefont {Luo}\ \emph {et~al.}(2015)\citenamefont {Luo},
  \citenamefont {Xiao}, \citenamefont {He}, \citenamefont {Sun},\ and\
  \citenamefont {Zhou}}]{luo2015photonic}%
  \BibitemOpen
  \bibfield  {author} {\bibinfo {author} {\bibfnamefont {W.}~\bibnamefont
  {Luo}}, \bibinfo {author} {\bibfnamefont {S.}~\bibnamefont {Xiao}}, \bibinfo
  {author} {\bibfnamefont {Q.}~\bibnamefont {He}}, \bibinfo {author}
  {\bibfnamefont {S.}~\bibnamefont {Sun}}, \ and\ \bibinfo {author}
  {\bibfnamefont {L.}~\bibnamefont {Zhou}},\ }\href@noop {} {\bibfield
  {journal} {\bibinfo  {journal} {Advanced Optical Materials}\ }\textbf
  {\bibinfo {volume} {3}},\ \bibinfo {pages} {1102} (\bibinfo {year}
  {2015})}\BibitemShut {NoStop}%
\bibitem [{\citenamefont {Brasselet}\ \emph {et~al.}(2013)\citenamefont
  {Brasselet}, \citenamefont {Gervinskas}, \citenamefont {Seniutinas},\ and\
  \citenamefont {Juodkazis}}]{brasselet2013topological}%
  \BibitemOpen
  \bibfield  {author} {\bibinfo {author} {\bibfnamefont {E.}~\bibnamefont
  {Brasselet}}, \bibinfo {author} {\bibfnamefont {G.}~\bibnamefont
  {Gervinskas}}, \bibinfo {author} {\bibfnamefont {G.}~\bibnamefont
  {Seniutinas}}, \ and\ \bibinfo {author} {\bibfnamefont {S.}~\bibnamefont
  {Juodkazis}},\ }\href@noop {} {\bibfield  {journal} {\bibinfo  {journal}
  {Physical review letters}\ }\textbf {\bibinfo {volume} {111}},\ \bibinfo
  {pages} {193901} (\bibinfo {year} {2013})}\BibitemShut {NoStop}%
\bibitem [{\citenamefont {Karimi}\ \emph {et~al.}(2014)\citenamefont {Karimi},
  \citenamefont {Schulz}, \citenamefont {De~Leon}, \citenamefont {Qassim},
  \citenamefont {Upham},\ and\ \citenamefont {Boyd}}]{karimi2014generating}%
  \BibitemOpen
  \bibfield  {author} {\bibinfo {author} {\bibfnamefont {E.}~\bibnamefont
  {Karimi}}, \bibinfo {author} {\bibfnamefont {S.~A.}\ \bibnamefont {Schulz}},
  \bibinfo {author} {\bibfnamefont {I.}~\bibnamefont {De~Leon}}, \bibinfo
  {author} {\bibfnamefont {H.}~\bibnamefont {Qassim}}, \bibinfo {author}
  {\bibfnamefont {J.}~\bibnamefont {Upham}}, \ and\ \bibinfo {author}
  {\bibfnamefont {R.~W.}\ \bibnamefont {Boyd}},\ }\href@noop {} {\bibfield
  {journal} {\bibinfo  {journal} {Light Sci. Appl}\ }\textbf {\bibinfo {volume}
  {3}},\ \bibinfo {pages} {e167} (\bibinfo {year} {2014})}\BibitemShut
  {NoStop}%
\bibitem [{\citenamefont {Osorio}\ \emph {et~al.}(2015)\citenamefont {Osorio},
  \citenamefont {Mohtashami},\ and\ \citenamefont {Koenderink}}]{osorio2015k}%
  \BibitemOpen
  \bibfield  {author} {\bibinfo {author} {\bibfnamefont {C.~I.}\ \bibnamefont
  {Osorio}}, \bibinfo {author} {\bibfnamefont {A.}~\bibnamefont {Mohtashami}},
  \ and\ \bibinfo {author} {\bibfnamefont {A.~F.}\ \bibnamefont {Koenderink}},\
  }\href@noop {} {\bibfield  {journal} {\bibinfo  {journal} {Scientific
  reports}\ }\textbf {\bibinfo {volume} {5}} (\bibinfo {year}
  {2015})}\BibitemShut {NoStop}%
\bibitem [{\citenamefont {Shitrit}\ \emph {et~al.}(2011)\citenamefont
  {Shitrit}, \citenamefont {Bretner}, \citenamefont {Gorodetski}, \citenamefont
  {Kleiner},\ and\ \citenamefont {Hasman}}]{shitrit2011optical}%
  \BibitemOpen
  \bibfield  {author} {\bibinfo {author} {\bibfnamefont {N.}~\bibnamefont
  {Shitrit}}, \bibinfo {author} {\bibfnamefont {I.}~\bibnamefont {Bretner}},
  \bibinfo {author} {\bibfnamefont {Y.}~\bibnamefont {Gorodetski}}, \bibinfo
  {author} {\bibfnamefont {V.}~\bibnamefont {Kleiner}}, \ and\ \bibinfo
  {author} {\bibfnamefont {E.}~\bibnamefont {Hasman}},\ }\href@noop {}
  {\bibfield  {journal} {\bibinfo  {journal} {Nano letters}\ }\textbf {\bibinfo
  {volume} {11}},\ \bibinfo {pages} {2038} (\bibinfo {year}
  {2011})}\BibitemShut {NoStop}%
\bibitem [{\citenamefont {Shitrit}\ \emph
  {et~al.}(2013{\natexlab{b}})\citenamefont {Shitrit}, \citenamefont {Maayani},
  \citenamefont {Veksler}, \citenamefont {Kleiner},\ and\ \citenamefont
  {Hasman}}]{shitrit2013rashba}%
  \BibitemOpen
  \bibfield  {author} {\bibinfo {author} {\bibfnamefont {N.}~\bibnamefont
  {Shitrit}}, \bibinfo {author} {\bibfnamefont {S.}~\bibnamefont {Maayani}},
  \bibinfo {author} {\bibfnamefont {D.}~\bibnamefont {Veksler}}, \bibinfo
  {author} {\bibfnamefont {V.}~\bibnamefont {Kleiner}}, \ and\ \bibinfo
  {author} {\bibfnamefont {E.}~\bibnamefont {Hasman}},\ }\href@noop {}
  {\bibfield  {journal} {\bibinfo  {journal} {Optics letters}\ }\textbf
  {\bibinfo {volume} {38}},\ \bibinfo {pages} {4358} (\bibinfo {year}
  {2013}{\natexlab{b}})}\BibitemShut {NoStop}%
\bibitem [{\citenamefont {Liu}\ \emph {et~al.}(2015)\citenamefont {Liu},
  \citenamefont {Ling}, \citenamefont {Yi}, \citenamefont {Zhou}, \citenamefont
  {Chen}, \citenamefont {Ke}, \citenamefont {Luo},\ and\ \citenamefont
  {Wen}}]{liu2015photonic}%
  \BibitemOpen
  \bibfield  {author} {\bibinfo {author} {\bibfnamefont {Y.}~\bibnamefont
  {Liu}}, \bibinfo {author} {\bibfnamefont {X.}~\bibnamefont {Ling}}, \bibinfo
  {author} {\bibfnamefont {X.}~\bibnamefont {Yi}}, \bibinfo {author}
  {\bibfnamefont {X.}~\bibnamefont {Zhou}}, \bibinfo {author} {\bibfnamefont
  {S.}~\bibnamefont {Chen}}, \bibinfo {author} {\bibfnamefont {Y.}~\bibnamefont
  {Ke}}, \bibinfo {author} {\bibfnamefont {H.}~\bibnamefont {Luo}}, \ and\
  \bibinfo {author} {\bibfnamefont {S.}~\bibnamefont {Wen}},\ }\href@noop {}
  {\bibfield  {journal} {\bibinfo  {journal} {Optics letters}\ }\textbf
  {\bibinfo {volume} {40}},\ \bibinfo {pages} {756} (\bibinfo {year}
  {2015})}\BibitemShut {NoStop}%
\bibitem [{\citenamefont {Ling}\ \emph {et~al.}(2015)\citenamefont {Ling},
  \citenamefont {Zhou}, \citenamefont {Yi}, \citenamefont {Shu}, \citenamefont
  {Liu}, \citenamefont {Chen}, \citenamefont {Luo}, \citenamefont {Wen},\ and\
  \citenamefont {Fan}}]{ling2015giant}%
  \BibitemOpen
  \bibfield  {author} {\bibinfo {author} {\bibfnamefont {X.}~\bibnamefont
  {Ling}}, \bibinfo {author} {\bibfnamefont {X.}~\bibnamefont {Zhou}}, \bibinfo
  {author} {\bibfnamefont {X.}~\bibnamefont {Yi}}, \bibinfo {author}
  {\bibfnamefont {W.}~\bibnamefont {Shu}}, \bibinfo {author} {\bibfnamefont
  {Y.}~\bibnamefont {Liu}}, \bibinfo {author} {\bibfnamefont {S.}~\bibnamefont
  {Chen}}, \bibinfo {author} {\bibfnamefont {H.}~\bibnamefont {Luo}}, \bibinfo
  {author} {\bibfnamefont {S.}~\bibnamefont {Wen}}, \ and\ \bibinfo {author}
  {\bibfnamefont {D.}~\bibnamefont {Fan}},\ }\href@noop {} {\bibfield
  {journal} {\bibinfo  {journal} {Light: Science \& Applications}\ }\textbf
  {\bibinfo {volume} {4}},\ \bibinfo {pages} {e290} (\bibinfo {year}
  {2015})}\BibitemShut {NoStop}%
\bibitem [{\citenamefont {Berry}(1984)}]{Berry45}%
  \BibitemOpen
  \bibfield  {author} {\bibinfo {author} {\bibfnamefont {M.~V.}\ \bibnamefont
  {Berry}},\ }\href {\doibase 10.1098/rspa.1984.0023} {\bibfield  {journal}
  {\bibinfo  {journal} {Proceedings of the Royal Society of London A:
  Mathematical, Physical and Engineering Sciences}\ }\textbf {\bibinfo {volume}
  {392}},\ \bibinfo {pages} {45} (\bibinfo {year} {1984})}\BibitemShut
  {NoStop}%
\bibitem [{\citenamefont {Kotlyar}\ \emph
  {et~al.}(2014{\natexlab{a}})\citenamefont {Kotlyar}, \citenamefont
  {Kovalev},\ and\ \citenamefont {Soifer}}]{kotlyar2014asymmetricModes}%
  \BibitemOpen
  \bibfield  {author} {\bibinfo {author} {\bibfnamefont {V.}~\bibnamefont
  {Kotlyar}}, \bibinfo {author} {\bibfnamefont {A.}~\bibnamefont {Kovalev}}, \
  and\ \bibinfo {author} {\bibfnamefont {V.}~\bibnamefont {Soifer}},\
  }\href@noop {} {\bibfield  {journal} {\bibinfo  {journal} {Optics Letters}\
  }\textbf {\bibinfo {volume} {39}},\ \bibinfo {pages} {2395} (\bibinfo {year}
  {2014}{\natexlab{a}})}\BibitemShut {NoStop}%
\bibitem [{\citenamefont {Kotlyar}\ \emph
  {et~al.}(2014{\natexlab{b}})\citenamefont {Kotlyar}, \citenamefont {Kovalev},
  \citenamefont {Skidanov},\ and\ \citenamefont
  {Soifer}}]{kotlyar2014asymmetricBeams}%
  \BibitemOpen
  \bibfield  {author} {\bibinfo {author} {\bibfnamefont {V.}~\bibnamefont
  {Kotlyar}}, \bibinfo {author} {\bibfnamefont {A.}~\bibnamefont {Kovalev}},
  \bibinfo {author} {\bibfnamefont {R.}~\bibnamefont {Skidanov}}, \ and\
  \bibinfo {author} {\bibfnamefont {V.}~\bibnamefont {Soifer}},\ }\href@noop {}
  {\bibfield  {journal} {\bibinfo  {journal} {JOSA A}\ }\textbf {\bibinfo
  {volume} {31}},\ \bibinfo {pages} {1977} (\bibinfo {year}
  {2014}{\natexlab{b}})}\BibitemShut {NoStop}%
\bibitem [{\citenamefont {Slobozhanyuk}\ \emph {et~al.}(2016)\citenamefont
  {Slobozhanyuk}, \citenamefont {Poddubny}, \citenamefont {Sinev},
  \citenamefont {Samusev}, \citenamefont {Yu}, \citenamefont {Kuznetsov},
  \citenamefont {Miroshnichenko},\ and\ \citenamefont
  {Kivshar}}]{slobozhanyuk2016enhanced}%
  \BibitemOpen
  \bibfield  {author} {\bibinfo {author} {\bibfnamefont {A.~P.}\ \bibnamefont
  {Slobozhanyuk}}, \bibinfo {author} {\bibfnamefont {A.~N.}\ \bibnamefont
  {Poddubny}}, \bibinfo {author} {\bibfnamefont {I.~S.}\ \bibnamefont {Sinev}},
  \bibinfo {author} {\bibfnamefont {A.~K.}\ \bibnamefont {Samusev}}, \bibinfo
  {author} {\bibfnamefont {Y.~F.}\ \bibnamefont {Yu}}, \bibinfo {author}
  {\bibfnamefont {A.~I.}\ \bibnamefont {Kuznetsov}}, \bibinfo {author}
  {\bibfnamefont {A.~E.}\ \bibnamefont {Miroshnichenko}}, \ and\ \bibinfo
  {author} {\bibfnamefont {Y.~S.}\ \bibnamefont {Kivshar}},\ }\href@noop {}
  {\bibfield  {journal} {\bibinfo  {journal} {Laser \& Photonics Reviews}\ }
  (\bibinfo {year} {2016})}\BibitemShut {NoStop}%
\bibitem [{\citenamefont {Mendoza-Hern{\'a}ndez}\ \emph
  {et~al.}(2015)\citenamefont {Mendoza-Hern{\'a}ndez}, \citenamefont
  {Arroyo-Carrasco}, \citenamefont {Iturbe-Castillo},\ and\ \citenamefont
  {Ch{\'a}vez-Cerda}}]{mendoza2015laguerre}%
  \BibitemOpen
  \bibfield  {author} {\bibinfo {author} {\bibfnamefont {J.}~\bibnamefont
  {Mendoza-Hern{\'a}ndez}}, \bibinfo {author} {\bibfnamefont {M.~L.}\
  \bibnamefont {Arroyo-Carrasco}}, \bibinfo {author} {\bibfnamefont {M.~D.}\
  \bibnamefont {Iturbe-Castillo}}, \ and\ \bibinfo {author} {\bibfnamefont
  {S.}~\bibnamefont {Ch{\'a}vez-Cerda}},\ }\href@noop {} {\bibfield  {journal}
  {\bibinfo  {journal} {Optics letters}\ }\textbf {\bibinfo {volume} {40}},\
  \bibinfo {pages} {3739} (\bibinfo {year} {2015})}\BibitemShut {NoStop}%
\bibitem [{\citenamefont {Saleh}\ and\ \citenamefont
  {Teich}(1991)}]{saleh1991fundamentals}%
  \BibitemOpen
  \bibfield  {author} {\bibinfo {author} {\bibfnamefont {B.~E.}\ \bibnamefont
  {Saleh}}\ and\ \bibinfo {author} {\bibfnamefont {M.~C.}\ \bibnamefont
  {Teich}},\ }\href@noop {} {\emph {\bibinfo {title} {Fundamentals of
  photonics}}}\ (\bibinfo  {publisher} {Wiley},\ \bibinfo {year}
  {1991})\BibitemShut {NoStop}%
\bibitem [{\citenamefont {Franke-Arnold}\ and\ \citenamefont
  {Jeffers}(2008)}]{franke2008orbital}%
  \BibitemOpen
  \bibfield  {author} {\bibinfo {author} {\bibfnamefont {S.}~\bibnamefont
  {Franke-Arnold}}\ and\ \bibinfo {author} {\bibfnamefont {J.}~\bibnamefont
  {Jeffers}},\ }\href@noop {} {\bibfield  {journal} {\bibinfo  {journal}
  {Structured Light and Its Applications, ed. Andrews, DL}\ ,\ \bibinfo {pages}
  {271}} (\bibinfo {year} {2008})}\BibitemShut {NoStop}%
\end{thebibliography}
\end{document}